# Lattice dynamics of the frustrated kagome compound Y-kapellasite

P. Doležal[1,2], T. Biesner[3], Y. Li[4,5], R. Mathew Roy[3], S. Roh[3], R. Valentí[4], M. Dressel[3], P. Puphal[6], A. Pustogow[2]

[1] *Charles University, Faculty of Mathematics and Physics, Department of Condensed Matter Physics, Ke Karlovu 5, 121 16 Prague 2, Czech Republic*
[2] *Institute of Solid State Physics, TU Wien, Vienna 1040, Austria*
[3] *1. Physikalisches Institut, Universität Stuttgart, 70569 Stuttgart, Germany*
[4] *Institut für Theoretische Physik, Goethe-Universität Frankfurt, 60438 Frankfurt am Main, Germany*
[5] *MOE Key Laboratory for Nonequilibrium Synthesis and Modulation of Condensed Matter, School of Physics, Xi'an Jiaotong University, Xi'an 710049, China*
[6] *Max Planck Institute for Solid State Research, 70569 Stuttgart, Germany*

**ABSTRACT**

Studying the magnetic ground states of frustrated antiferromagnets provides unique insight into the stability of quantum spin liquids, even if the anticipated state is not realized towards $T = 0$. Particularly relevant are structural modifications setting in at temperatures where the magnetic correlations come into play. Here we explore the lattice dynamics of Y-kapellasite ($Y_3Cu_9(OH)_{19}Cl_8$) single crystals by infrared spectroscopy in combination with ab initio calculations. We observe significant changes in the phonon spectra at $T_s = 32$ K, that gradually evolve down to low temperatures. The increase in the number of phonon modes provides evidence for a lowering of symmetry and we discuss several possibilities of crystal structure modifications. Our analysis also reveals that the structural variation involves exclusively H and O atoms, while the other atoms remain rather unaffected. An 8% red shift of the lowest-lying phonon mode upon cooling indicates strong magneto-elastic effects upon decoupling Cu-6f hexagons through the lattice vibrations.

## I. INTRODUCTION

The search for quantum spin liquids (QSL) remains an intensely investigated topic in solid-state physics [1,2]. Hallmark is the absence of magnetic order down to the lowest temperatures, along with long-range entanglement of fluctuating spin excitations. Magnetic frustration, which prevents long-range antiferromagnetic (AFM) order, is a way of promoting such a state of matter. Several crystal lattices are close to these conditions, such as the kagome lattice with nearest neighbor antiferromagnetic interactions. An intensely discussed candidate material is herbertsmithite which consists of a close-to ideal kagome lattice of spin-1/2 Cu atoms separated by interlayer Zn atoms [3]. A small but significant complication in this material is a partial mixing of Zn and Cu atoms on the same Wyckoff position creating structural disorder and impurities in the lattice [4-6]. Apart from geometrical frustration, disorder has been suggested to play a major role in QSL candidates [4-13]. In the case of herbertsmithite this issue has been tackled by substituting interlayer Zn atoms with bigger elements that prevent the formation of antisite Cu defects. This resulted in intense research endeavors on the new family of kagome copper hydroxides [14]. One prominent member is the Y substitution having a crystal structure similar to kapellasite [15] but with two different polymorphs. $YCu_3(OH)_6Cl_3$ was prepared first, but its stability is sensitive to moisture [16,17]. A distinct way of preparation leads to the stable polymorph $Y_3Cu_9(OH)_{19}Cl_8$, which allows the growth of large single crystals used in this study [18,19].

$Y_3Cu_9(OH)_{19}Cl_8$ is a Mott insulator with charge-transfer gap of 3.6 eV [20] whose magnetic ground state has been debated over the past years. First, the long-range antiferromagnetic order with $T_N = 2.2$ K [18] was reported, then based on muon spin relaxation experiments on powder samples the ground state seemed to exhibit persistent dynamics [17], which was recently also found for the new sister compound $YCu_3(OH)_{6.5}Br_{2.5}$ [21]. Some experimental studies also proposed the origin of the 2.2 K anomaly to be a sign of CuO impurities [22], but more recently $^1$H and $^{35}$Cl NMR studies [19,23] on high quality $Y_3Cu_9(OH)_{19}Cl_8$ single crystals proved the existence of long-range magnetic order of $Q = (1/3 \times 1/3)$ type [24] with the order parameter growing in a mean-field fashion below $T_N = 2.2$ K [19,23].

The crystal lattice is rhombohedral $R$-3 (148) with the hexagonal lattice parameters $a_{hex} = 11.5350(8)$ Å, $c_{hex} = 17.2148(12)$ Å [18]. The Cu atoms create a slightly distorted kagome lattice but the full determination of the crystal structure was also not unique in the past [17-19]. The reason is the presence of H atoms that are difficult to detect in X-ray diffraction and usually the determination is based on additional and more indirect assumptions. Further neutron diffraction experiments on single crystalline samples, however, were able to detect at $T_s = 32$ K an abrupt change in intensity of the (6 0 0) and (0 0 18) diffraction maxima, but no particular modification of the crystal lattice could be resolved [19]. This transition at 32 K was also observed as an anomaly in specific heat and, even more pronounced, in the thermal expansion coefficient [19]. At the same time, an onset of magnetic correlations was observed in susceptibility and NMR spin-lattice relaxation rate [23,25]. These findings point out the relevance of the structural modifications for the magnetic ground state, but it remains an open question whether the anomaly has its origin purely in the crystal structure and what exactly is happening below $T_s = 32$ K. This motivates our present study of lattice dynamics by infrared (IR) spectroscopy in comparison with Density Functional Theory (DFT) calculations of the phonon modes, which are sensitive to structural modifications and magneto-elastic coupling effects. The article is structured in to five chapters: Introduction, Experimental methods and sample characterization, Optical analysis, Theoretical calculations, Discussion, and Conclusion.

## II. EXPERIMENTAL METHODS AND SAMPLE CHARACTERISATION

The crystal structure model used in this study was determined from single crystal X-ray diffraction in [18], see Fig. 1. Each H atom is bonded to one O with the exception of the O between the kagome layers, see the dotted ellipse in Fig. 1. Here the six indicated H atoms are bonded to one O but with occupancy 1/6 simulating the disorder of this H atom. The neutron powder diffraction on the deuterated sample confirmed this structural model but with the absence of H or D atoms bonded to the O between the kagome layers leading to the stoichiometry of $Y_3Cu_9(OH)_{18}OCl_8$ [17]. Further experiments for determining the stoichiometry of this compound led to $Y_{3.00(3)}Cu_{9.06(9)}O_{19.0(2)}H_{18.7(3)}Cl_8$ which rather suggests the first structural model [18,19]. In our optical spectra in Figs. 2 and 3 we observe hydrogen stretching modes at 3280 and 3400 cm$^{-1}$ along both directions. A third branch at 3500 cm$^{-1}$, which is close to the resonance frequency of free OH ions, is visible only for electric field polarized along the ab-plane suggesting that this is related to the inter-plane OH bonds, which further corroborates the existence of these H-sites.

Single crystals of $Y_3Cu_9(OH)_{19}Cl_8$ were prepared by hydrothermal synthesis methods as described in our previous works [18,19]. The crystal structure of the studied compound $Y_3Cu_9(OH)_{19}Cl_8$ was checked on a small sample from the same batch as the studied samples confirming the same structural model as in [18]. The lattice parameters were also directly determined on the studied sample by measurement of (00l) and (hh0) diffraction maxima. The resulting lattice parameters $a = 11.5554(5)$ Å, $c = 17.2223(4)$ Å were determined from the Cohen-Wagner plot (see Fig. S1 in [26]) and are in good agreement with those from the single crystal X-ray diffraction. The symmetrical $\theta$-$\theta$ scan proves that the amount of impurities is below the detection limit of X-ray diffraction.

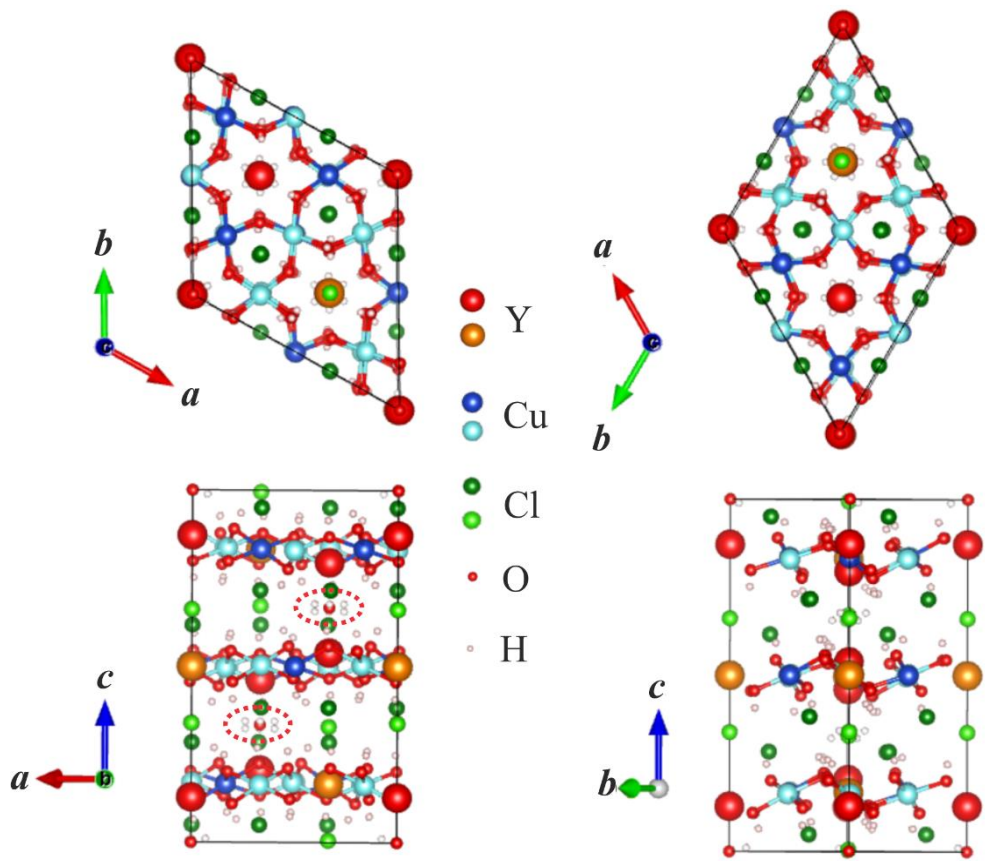

Fig.1: Crystal structure of $Y_3Cu_9(OH)_{19}Cl_8$ as determined in Ref. [18]. The unit cell is viewed along the *c* direction (top) as well as from the perpendicular (side) view (bottom). The dashed ellipse marks the position of O atoms in 1a Wyckoff's position between the kagome layers. This O atom is surrounded by six H atoms, whose site occupancy is 1/6, corresponding to the H disorder around these O atoms.

Optical conductivity was calculated from the measured reflectivity (Fig. S2, Fig. S3), recorded in the range $40 - 48,000$ cm$^{-1}$, using the Kramers-Kronig relations. For that, the data were extrapolated by a constant function and by a smooth reflectance at low and high frequencies, respectively. The X-ray reflectance was calculated in the range $8 \cdot 10^5 - 2.4 \cdot 10^8$ cm$^{-1}$ by the

XRO.exe software [27]. The temperature-dependent optical reflectivity was measured at normal incidence in the far-infrared (10 – 700 cm$^{-1}$, Hg lamp as the source and bolometer detector), mid-infrared (500 – 8,000 cm$^{-1}$, Globar and MCT detector), near-infrared (2,000 – 12,000 cm$^{-1}$, W lamp and InSb detector) and visible (2,000 – 20,000 cm$^{-1}$, W lamp and Si-diode detector) ranges using FTIR spectrometers (Bruker 113v, Vertex 80v) and associated low-temperature cryostats. These measurements were complemented by THz data at low frequencies (7 – 50 cm$^{-1}$, time-domain THz from Ref. [25]) as well as high-frequency data up to the ultraviolet (6,000 – 48,000 cm$^{-1}$, ellipsometry from Ref. [20]). Temperature dependence of reflectance was measured on three samples: Sample 1 – (far-infrared) and Sample 2 – (mid-infrared) at 3, 10, 20, 50, 100, 150, 200, 295 K and for Sample 3 – (mid-infrared) with fine temperature steps between 16 and 50 K.

## III. OPTICAL ANALYSIS

The optical properties are characterized by the dielectric tensor, which is symmetric. Together with its rhombohedral symmetry the compound possesses only one dielectric axis. This is also valid for the complex dielectric tensor [28] and means that for the determination of optical properties, it is necessary to measure reflectance with electrical field intensity $E$ parallel and perpendicular to the dielectric axis in our case parallel and perpendicular to the $c$-axis of the hexagonal unit cell i.e. $E \parallel c$ and $E \parallel ab$. The temperature dependence of the measured optical conductivity is shown in Fig. 2 and Fig. 3 for both polarizations. Our study focuses on the lattice dynamics and therefore only far and mid-IR regions are shown. We can observe the typical temperature dependence when the spectral weight of phonon modes increases as the anharmonicity of the potential is less pronounced. Such behavior is observed during cooling down to 100 K in both polarizations. Another typical property is the shift of phonon modes to higher energies (blue shift) as the temperature is reduced. This is the effect of the contraction of the inter-atomic distances upon cooling. Still, we have to be careful in less symmetrical unit cells, where one of the lattice parameters can have an opposite length change than the volume and also when the atoms are in a free Wyckoff position.

In Y-kapellasite, however, besides the pronounced anomaly at 32 K, dilatometry experiments [19] did not reveal negative thermal expansion – which does not rule out that some interatomic distances shorten while others become longer, of course. The opposite shift to lower frequencies (red shift) is a sign of additional effects, for instance magneto-elastic coupling [20,29] or anomalous temperature dependence of interatomic distances. We observe a shift to lower frequencies in the phonon modes at 80, 650, and 3270 cm$^{-1}$ in $E \parallel ab$ (Fig. 2) or at 3270 cm$^{-1}$ in $E \parallel c$ (Fig. 3). The relative red shift of almost 8 % of the resonance frequency of the lowest-lying phonon feature at 80 cm$^{-1}$ is especially large in comparison with the other mentioned modes, see the detail and positions of the peak in Fig. 4a) and b). Note, the phonon frequency is comparable to the exchange energy and $\Theta_w$ of 100 K = 69.5 cm$^{-1}$ of Y-kapellasite [18], suggesting that the modes are particularly susceptible to magneto-elastic coupling. A closer view on the spectral weight profile in Fig. 4c) indicates contributions of multiple phonon modes and a change of its shape below 50 K.

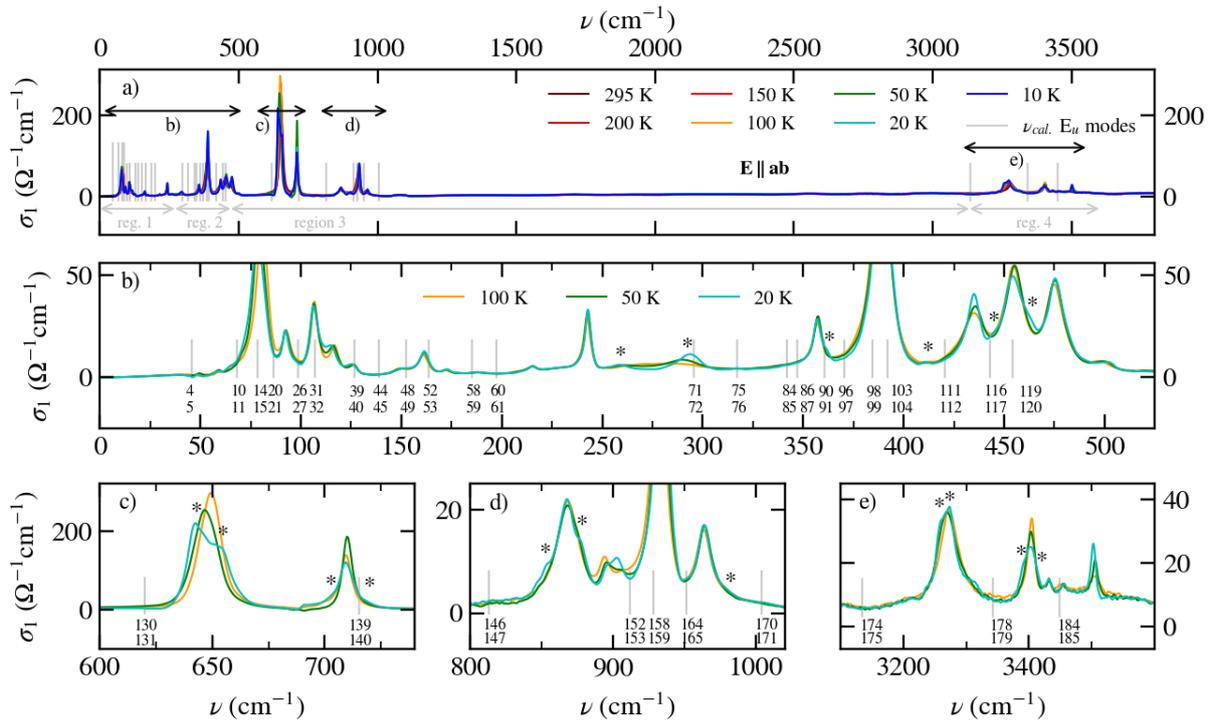

**Fig. 2:** a) Optical conductivity ($E \parallel ab$) in far-IR (Sample 1) and mid-IR (Sample 2) regions at various temperatures. b), c), d), and e) panels show selected regions in detail. The vertical grey line segments represent the calculated energy of the phonon modes. The calculated modes are labeled in increasing order as the energy of modes increases. We note that $E_g$, $A_u$, and $A_g$ modes are not shown therefore these numbers are missing below the vertical grey lines. The stars represent the appearance of new phonon modes.

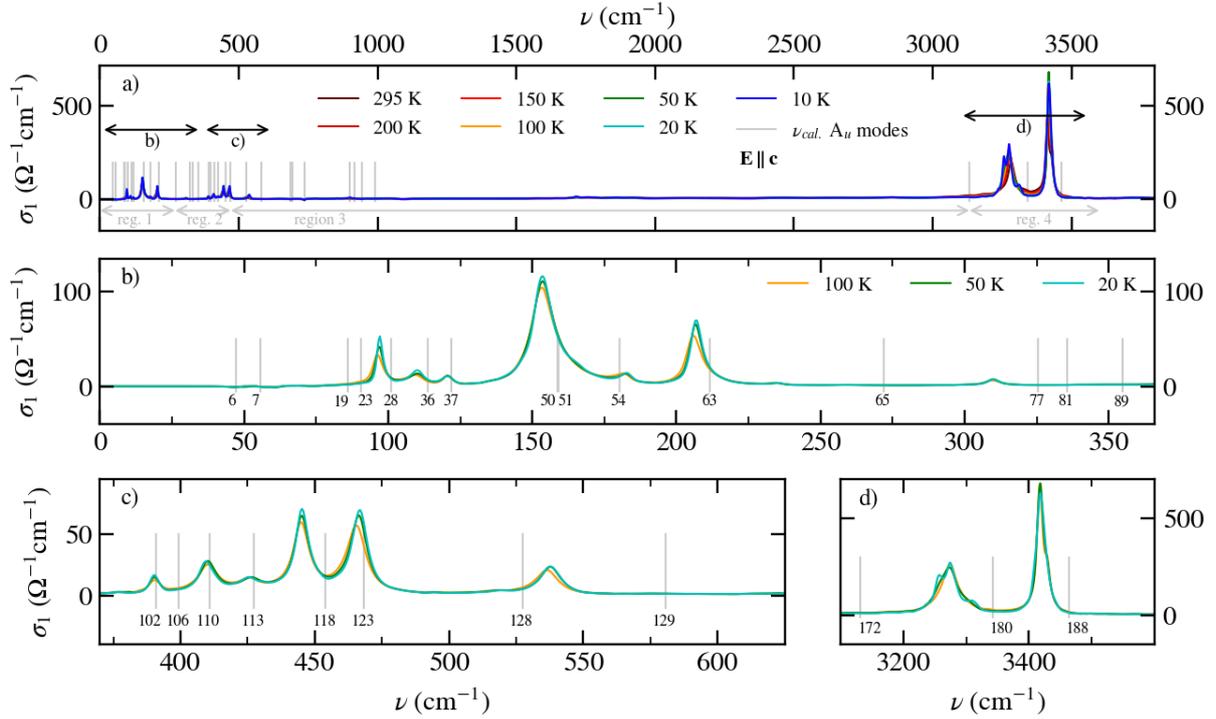

**Fig. 3:** a) Optical conductivity ($E \parallel c$) in far-IR (Sample 1) and mid-IR (Sample 2) regions at various temperatures. b), c), d), and e) panels show selected regions in detail. The vertical grey line segments represent the calculated energy of the phonon modes. The calculated modes are labeled in increasing order as the energy of modes increases. We note that $E_g$, $E_u$, and $A_g$ modes are not shown therefore these numbers are missing below the vertical grey lines.

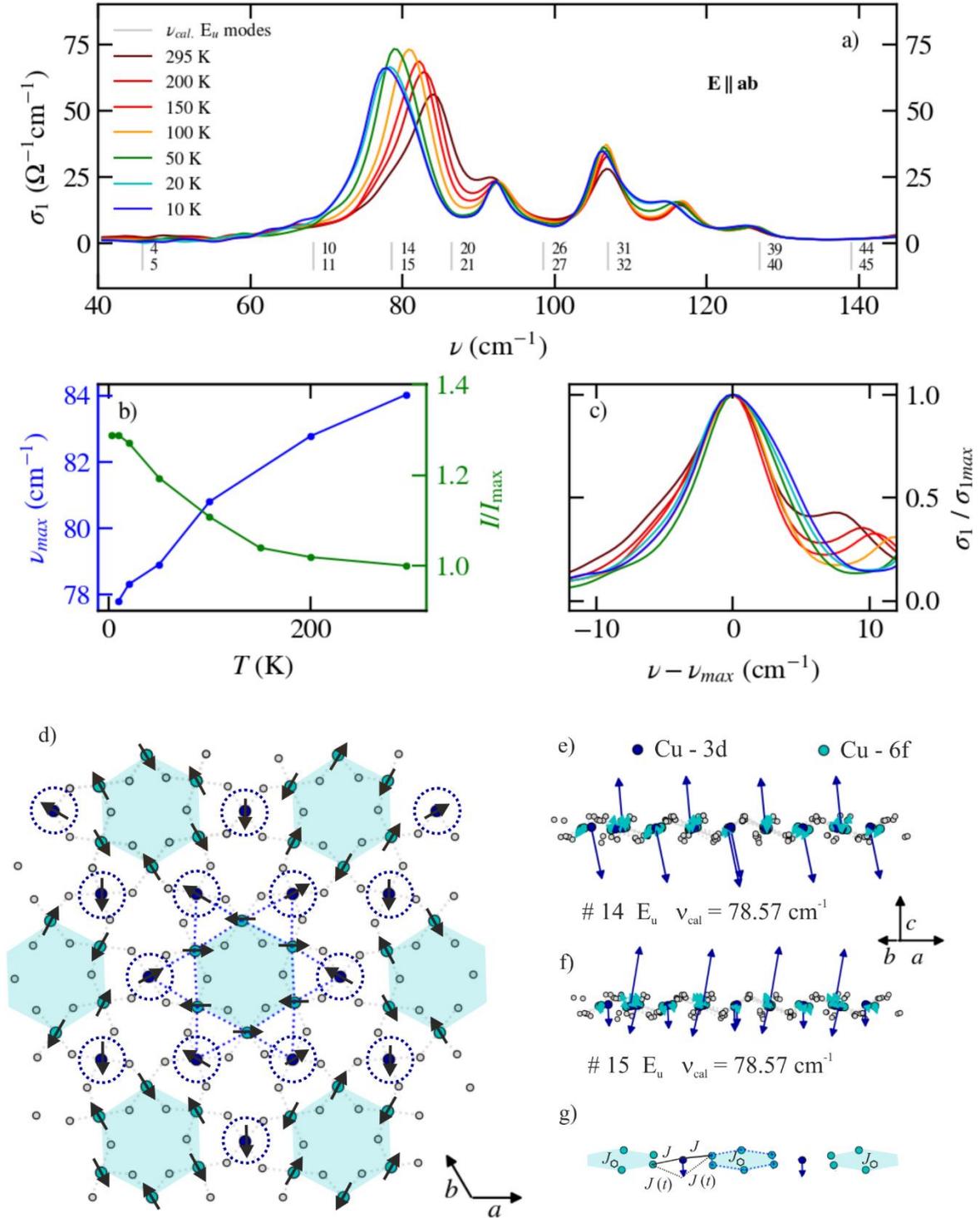

**Fig. 4:** a) Temperature evolution of spectral weight of lowest frequency phonon modes around 80 - 120 cm$^{-1}$. b) Temperature dependence of the peak position and spectral weight around 80 cm$^{-1}$ from panel a) reveals pronounced red shifting upon cooling. c) Comparison of the shape profile of these maxima, changing their asymmetry below 50 K. d-f) Motion of Cu atoms in modes 14 and 15 involves strong motion of the Cu-3d positions (dark blue) between the hexagons, while Cu-6f motions within the hexagons (cyan) are much weaker. g) Illustration of how the strong motion of the Cu-3d atoms influences the change of $J$, while $J_{\hexagon}$ within the hexagons remains rather unaffected by the vibration.

There are also phonon modes that shift to a higher frequency during cooling down to 50 K, while, there is an opposite trend at lower temperatures. This is observed for example for the modes at 388, 434, 453, 709 cm$^{-1}$ in $E \parallel ab$ (Fig. 2) or 410 cm$^{-1}$ in $E \parallel c$ (Fig. 3). We can also observe the appearance of new vibrational modes with very small spectral weight in $E \parallel ab$ polarization below 50 K. These new modes appear around the already existing phonon modes, which exhibit the shift to lower frequencies accompanied by a decrease in spectral weight. The lowest new peak appearance can be identified around 275 cm$^{-1}$, but in this case the new peaks appear already between 50 and 100 K, the original mode between them is very weak. The appearance of new peaks was not observed in the polarization $E \parallel c$ with one exception of the mode at 3270 cm$^{-1}$.

To determine the origin of these new phonon modes, we measured the temperature dependence of the reflectance in fine temperature steps on sample 3. The corresponding optical conductivity is shown in Fig. 5, displaying the mode at 650 cm$^{-1}$ for $E \parallel ab$ and the modes in the interval 3200 – 3400 cm$^{-1}$ for $E \parallel c$, because the pronounced changes in the conductivity spectra are at these frequencies. The phonon mode at 650 cm$^{-1}$ in $E \parallel ab$ is slightly asymmetric already at high temperatures. For simplicity the feature is fitted by two profiles using the Lorentz oscillator model to describe the asymmetry. During cooling, the spectral weight decreases and two new modes appear on the sides of the previous mode. In total we have four Lorentz oscillators at 16 K. In Fig. 5a) it is visible that this change is of continuous character starting around 32 K. At low temperatures the spectral profile can be fitted by three modes instead of one at 50 K. In the $E \parallel c$ polarization, the new modes appear only in the region 3200 – 3400 cm$^{-1}$ between 36 and 32 K, which confirms that these changes are of the same origin, related to the structural anomaly at $T_s = 32$ K [19].

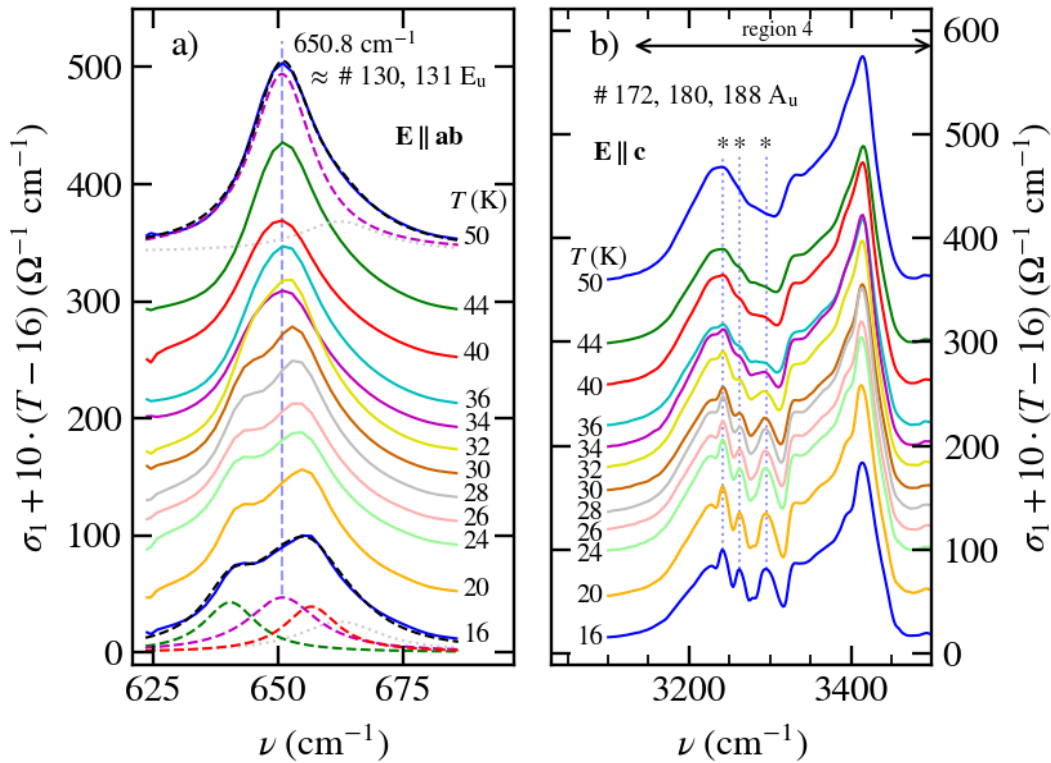

**Fig. 5:** a) Detail of optical conductivity around 650 cm$^{-1}$ measured on the Sample 3 in polarization $E \parallel ab$. The dashed lines correspond to fits of phonon modes by the Lorentz oscillator model. The grey dotted line was used for the simulation of the mode asymmetry. b) Optical conductivity of the Sample 3 in polarization $E \parallel c$ showing the appearance of new phonon modes on the positions marked by stars. a) and b) the spectra are shifted (using a constant 10 x (T - 16)) along the y-axis for clarity. The numbers of modes represent the modes whose energy should be in the displayed region based on the theoretical calculations.

## IV. THEORETICAL CALCULATIONS

To investigate the lattice dynamics, we performed density functional theory calculations starting from the structural model in Ref. [18] at 173 K. The disorder of H atoms (fractional occupancy) around the O between Kagome layers in this structure was modeled using the virtual crystal approximation (VCA) while maintaining crystal symmetry. The phonon frequencies were calculated using the PHONOPY package [30,31] by diagonalization of the dynamical matrices constructed from the force constants. The force constants were determined by the Vienna ab initio package (VASP) with the Perdew Burke-Ernzerhof parameterization of the generalized gradient approximation [32]. We adopted a cutoff energy of 520 eV and Monkhorst-pack k-points generated with 4 × 4 × 4. The symmetries or irreducible representations (Au, Eu, Ag, Eg) of each phonon mode were determined by the behavior of displacements of atoms under the symmetry operations. The character of each class was calculated for atom displacements of each phonon mode and compared with the character table of -3 point group leading to the finding of irreducible representation.

It is useful to compare the activity of atoms in calculated phonon modes which helps to understand the observed changes in the previous chapter. For that purpose, we plot the average displacement of the atoms in the corresponding Wyckoff position, see Fig. 6. We especially stress that the displacement from equilibrium does not have to be the same for all atoms at one Wyckoff position, but we are using these sets of Wyckoff positions for simplicity and comparison. The phonon modes can be divided into four groups marked in Fig. 6b and as grey arrows in Figs. 2 and 3: regions 1, 2, 3, and 4 in Fig. 6. In region 1 the displacement of Cu, Cl, and Y atoms is dominant, while in region 2 Cu and O atoms move prominently and Y and Cl are more stable. Also in region 2 the movement of H atoms, which are disordered around O between layers (H-6f1), is significant. Region 3 is fully occupied by the movement of all H atoms while the activity of the rest of the atoms starts can eventually be neglected. The highest-frequency region 4 exhibits almost exclusively O-H stretching vibrations. It is possible to observe here three sub-regions dominated by H-6f3, H-6f2, and H-6f4 atoms as the energy increases, see Fig. 6d,e. The activity of H-6f1 atoms is very small in region 4.

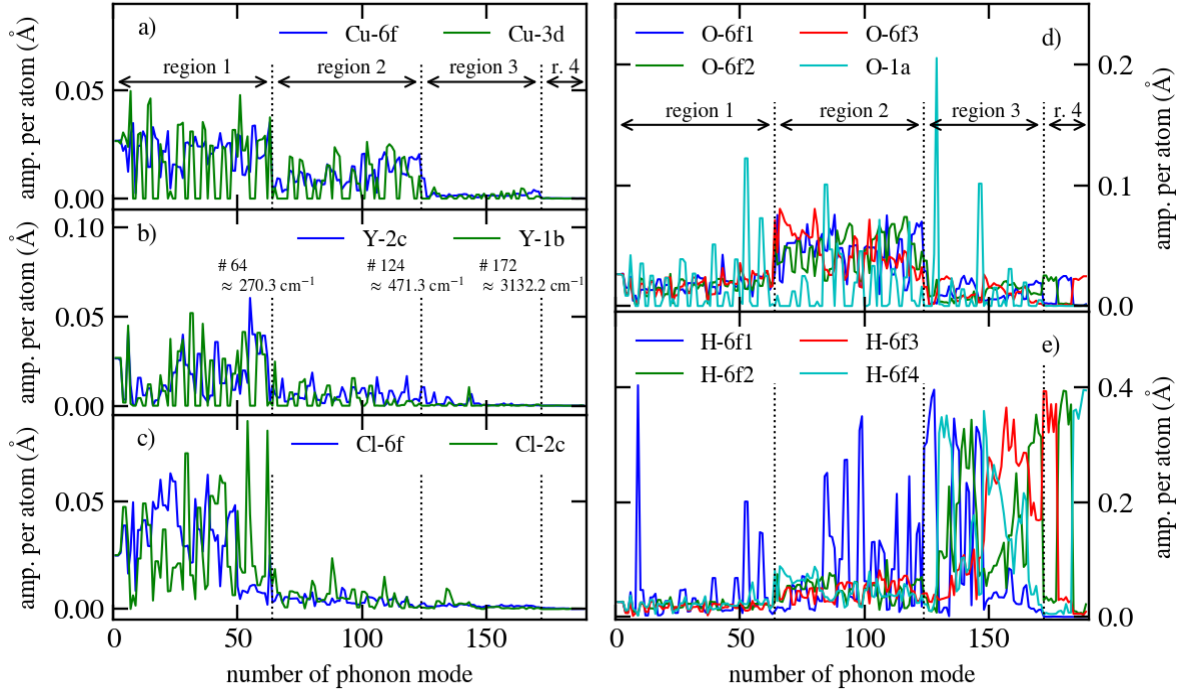

**Fig. 6:** Amplitude of the displacement for all calculated phonon modes. The dotted vertical lines mark the boundaries of the regions. a), b), c), d), and e) panels show displacement of Cu, Y, Cl, O, and H atoms, respectively.

## IV. DISCUSSION

Optical spectroscopy observes mainly phonon modes in the vicinity of the Γ point, because the wave vector of the light is very small compared to the size of the Brillouin zone. Therefore we only need to know the small group (group of the $k$ vector) at $k = 0$, which is always isomorphic to some of the point groups. In our case, we have $R$-3 (148) space group, with a point group - 3. The symmetries of phonon modes are labeled according to the irreducible representation of mechanical representation, which is a direct product of equivalence representation and representation of the polar vector:

$$(14\ A_g + 11\ A_u + 10\ ^1E_g + 10\ ^2E_g + 9\ ^1E_u + 9\ ^2E_u) \times (A_u + {}^1E_u + {}^2E_u) =$$

$$29\ A_g + 34\ A_u + 34\ ^1E_u + 34\ ^2E_u + 29\ ^1E_g + 29\ ^2E_g = 29\ A_g + 34\ A_u + 34\ E_u + 29\ E_g. \qquad (1)$$

We note that the representations labeled by E are two-dimensional, but $^1E$ and $^2E$ are one-dimensional. In total, we have 189 phonon modes, which is in agreement with 63 atoms in the primitive rhombohedral unit cell. However, only the anti-symmetric optically active phonon modes have non-zero dipole moment, which reduces the modes to 34 $A_u$ + 34 $E_u$. The symmetries of acoustic phonon modes are $A_u + E_u$, which finally reduced to 33 $A_u$ + 33 $E_u$ modes observable in our measurements. The lattice vectors of the primitive unit cell are not orthogonal, therefore the displacement of the atoms in the phonon modes $A_u$ and $E_u$ are also not orthogonal, but only $A_u$ modes have non-zero dipole moment projection to the $c_{hex}$ lattice vector. The opposite is valid for $E_u$ modes which only have a non-zero projection to $\mathbf{a_{hex}b_{hex}}$ plane. This allows the separation of $A_u$ and $E_u$ modes, which is crucial for the interpretation of results. The ideal situation is the ascription of calculated modes to experimental ones. This is usually straightforward in the case with only a few atoms in the primitive unit cell, but it becomes complicated in our case, because not all modes have large enough spectral weight to observe them. Last and most important is the H disorder on O atoms between the kagome layers. The formula unit contains 58 atoms, but the structural model includes 63 atoms, because of H disorder. Strictly speaking, this also means that, the number of phonon modes is not well determined. There are 174 modes considering 58 atoms and 189 modes with 63 atoms. These are all the reasons why we cannot provide a complete assignment each mode. Instead, we will focus on the temperature dependences in the following discussion.

The appearance of small peaks in the optical conductivity within the temperature range 32 – 36 K is in good agreement with the temperature of the anomaly in specific heat and thermal expansion, which was ascribed to structural modifications [19]. In the following, we will discuss possibilities for the open question of the type of structural modification. For clarity, we discuss three possible scenarios. Firstly, there can be a reduction in the lattice symmetry, like a lattice distortion. The rhombohedral crystal lattice would become monoclinic after the distortion. Which means that such a point group does not have two-dimensional irreducible representations and all the $E_u$ modes should split in two. This transition is sometimes continuous and connected with the formation of domains causing the mentioned splitting. The second possibility is the decrease of the point crystal symmetry without significant lattice distortion. This increases the volume of the primitive unit cell and leads to the appearance of the new phonon modes in the spectra; it will not look like a continuous splitting of the modes. The third possibility are structural modifications like an abrupt change occurring only in free Wyckoff positions, which in principle doesn't have to lead to the decrease of the symmetry. The number of phonon modes then remains unchanged and only a shift of the modes will be observed. In Fig. 2, we see the appearance of faint new modes at low temperatures, but the high-temperature modes remain present. The spectral weight of new modes increases continuously suggesting the continuous transition (see Fig. S4b), but in frequency new modes are distinct, which makes the first option less probable. The third option can be ruled out because the number of phonon modes increases. So the second scenario should possibly suit our case. There is another interesting aspect: the new peaks appear only above 350 cm$^{-1}$, which shows that the changes of Y, Cu, and Cl in the crystal structure have to be very small if they exist, see the regions in Fig. 6. Therefore the crystal structure symmetry seems to be lowered namely by hydrogen and oxygen atoms.

As mentioned above, the assignment of phonon modes is very complicated with one exception the phonon mode at 650 cm$^{-1}$, which can be assigned to mode numbers 130, and 131 moreover this phonon mode has one of the most pronounced changes in the spectra in $\boldsymbol{E \parallel ab}$ polarization. The results of theoretical calculations give us the atomic displacement of H and O atoms for this phonon mode. They are shown in Fig. 7 in one kagome layer (equivalent way how to display the displacement in the primitive unit cell). It is well seen that the motion is dominated by H-6f$_4$ atoms and is located namely within the kagome layer. It could seem that structural modifications are related only to the basal plane (kagome layer), but this is not fully true if we look at A$_u$ modes at region 4 in Fig. 3d or Fig. 5b, where we also see significant modifications of the spectra and the motion of H atoms is mainly perpendicular to the basal plane, see modes # 172, # 180, # 188 in Fig. S5, Fig. S6, Fig. S7 in [26], respectively.

The leading role of H atoms in structural modification can be easily understood if we assume that H is only bound to the O atom and is not a part of other bonds. It can then be the sensitive probe for other interactions than other more coordinated and heavy atoms. One of such interaction can be short-range magnetic correlations, as the $^1$H NMR results suggest. This could be verified by infrared spectroscopy in very high magnetic fields, because our measurements up to 7 T didn't show any noticeable change, see Fig. S8.

Finally we focus on magneto-elastic coupling which was noted in previous works on Y-kapellasite [25] and the related kagome system herbertsmithite [29,33]. Note that this does not necessarily have to be linked to the structural transition at 32 K, as the latter is dominated by hydrogen and oxygen sites while the Cu$^{2+}$ kagome plane remains rather unaffected. Magnetic interactions coupled to the lattice are a natural source for the red shift of phonon modes, as we mentioned above. The reason is that if the atomic displacement upon the vibration yields a lower total energy, the lattice sites stay longer in this configuration. As a result, the phonon resonance is shifted to lower frequencies. Since the most pronounced effect is expected when the energies are comparable, for $J \approx \Theta_w = -100$ K of Y-kapellasite [18] this corresponds to a frequency around 70 cm$^{-1}$. This is in agreement with the large red shift of the feature around 80 cm$^{-1}$, see Fig. 4. While the assignment of the measured peak in the spectrum to the calculated modes is not unambiguous in our case, from their resonance frequencies matching the experimentally observed feature we infer that the modes number 14, 15, and possibly also 10, 11 are involved (see atomic displacements of these modes in Figs. S9 – S12 in [26]). Quite exceptional in these modes is the displacement of Cu atoms – particularly Cu-3d movements, that are sitting between the Cu-6f hexagons – which is almost oriented perpendicular to the kagome layers and thus strongly impacts the antiferromagnetic coupling between Cu-6f hexagons. The Cu-atomic motions related to modes 14 and 15, illustrated in Fig. 4 d-f), reveal a strong amplitude of the Cu-3d sites that connect the Cu-6f hexagons, whereas the Cu-6f sites within the hexagons are only weakly displaced with respect to each other. This reminds strongly of the unusual $Q = (1/3 \times 1/3)$ antiferromagnetic order setting in below $T_N = 2.2$ K with a phase shift of $2\pi/3$ between the hexagons that are coupled via $J \sim J_\circ$, which has been consistently evidenced by theory [24] and experimentally via $^{35}$Cl [19] and $^1$H NMR [23]. We thus infer that the pronounced non-thermal behavior of the lowest-frequency phonons results from interrupting the exchange interactions between the hexagons (formed of Cu-6f sites, see Fig. 4d), within which the spins tend towards antiferromagnetic alignment. Without the interconnection via the Cu-3d sites the Cu-6f hexagons become rather uncoupled. This way, suppression of long-range antiferromagnetic correlations by atomic displacements causes pronounced magneto-elastic coupling and, thus, strongly affects these lowest-lying phonon modes.

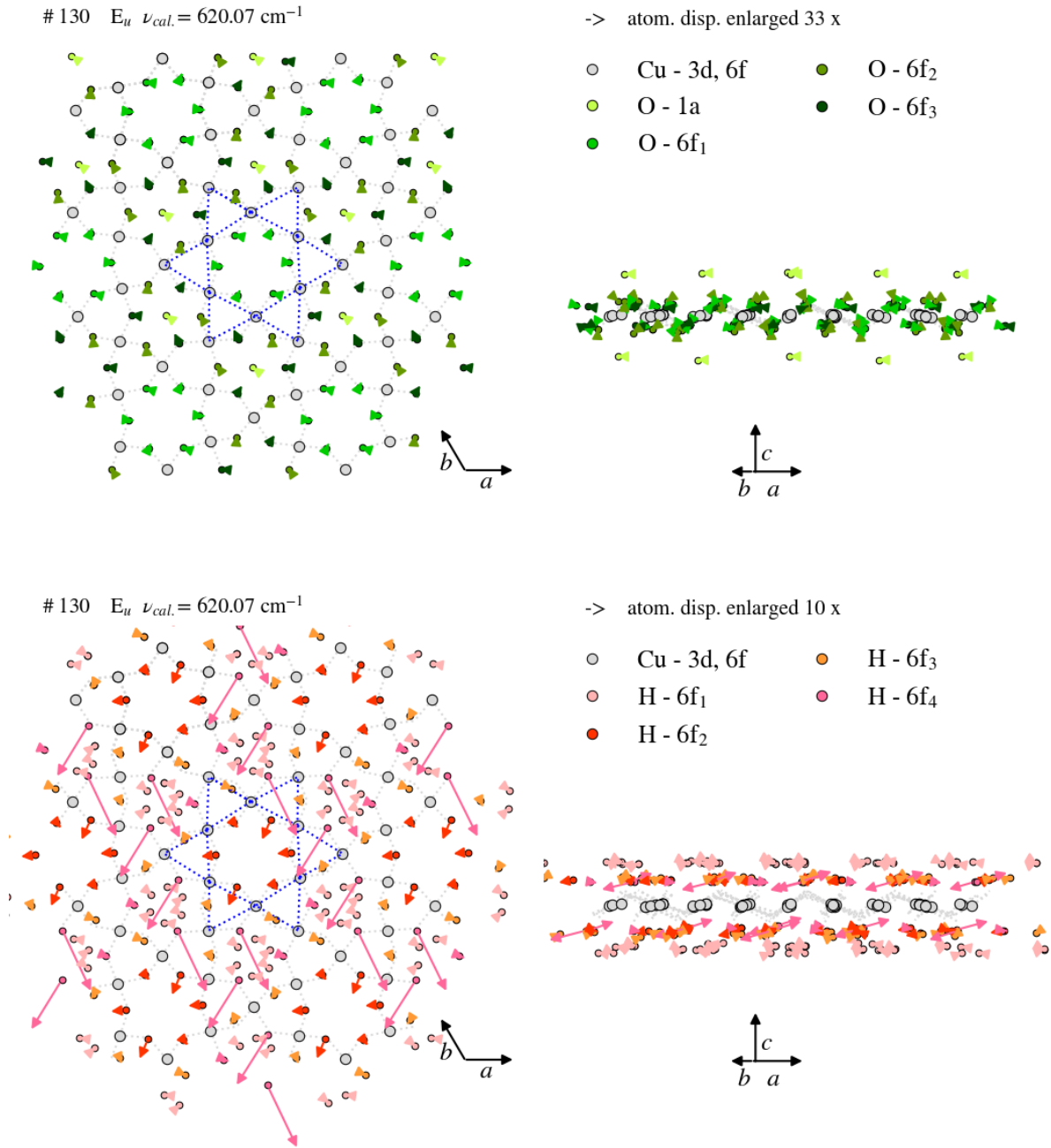

**Fig. 7:** Atomic displacement in Eu phonon mode # 130, the result of the theoretical calculation. We note that the displacement is enlarged for better visibility. The displacement of mode # 131 is shown in Fig. S13 in [26].

## V. CONCLUSION

In conclusion, our experimental and theoretical investigations confirm the structural origin of the changes in the optical conductivity spectra of the $Y_3Cu_9(OH)_{19}Cl_8$ compound. We found that the crystal structure modifications involve almost exclusively H and O atoms, which are also responsible for the lowering of the whole crystal structure as the other atoms remain rather unchanged. This is in agreement with the previous structural studies, which were unable to provide a more detailed description of the crystal structure changes. Our results also rule out that the structural modification is only the ordering of disordered H atoms around O atoms between kagome layers. Moreover, our calculations provide direct insight into the disturbance of antiferromagnetic interactions between Cu-6f hexagons by strong out-of-plane motion of Cu-3d atoms, which expresses in heavily pronounced magneto-elastic coupling of this lowest-frequency phonon feature, that exhibits almost 8% red shift upon cooling from room temperature towards $T \to 0$. We also suggest that short-range magnetic correlations, felt by the O atoms on the Cu-O-Cu superexchange path, can play the leading role in the movement of H atoms and lead to the lowering of crystal symmetry.

## ACKNOWLEDGEMENTS

This work was supported by the Deutsche Forschungsgemeinschaft (DFG) for financial support via DR228/68-1 and via TRR288 – 422213477 and by the Czech Science Foundation via research project GAČR 23-06810O. A.P. acknowledges support by Hochschuljubiläumsfonds der Stadt Wien (H-918729/2022).

Supplemental material for:

# Lattice dynamics of the frustrated kagome compound Y-kapellasite

P. Doležal[1,2], T. Biesner[3], Y. Li[4,5], R. Mathew Roy[3], S. Roh[3], R. Valentí[4], M. Dressel[3], P. Puphal[6], A. Pustogow[2]

[1] Charles University, Faculty of Mathematics and Physics, Department of Condensed Matter Physics, Ke Karlovu 5, 121 16 Prague 2, Czech Republic
[2] Institute of Solid State Physics, TU Wien, Vienna 1040, Austria
[3] 1. Physikalisches Institut, Universität Stuttgart, 70569 Stuttgart, Germany
[4] Institut für Theoretische Physik, Goethe-Universität Frankfurt, 60438 Frankfurt am Main, Germany
[5] MOE Key Laboratory for Nonequilibrium Synthesis and Modulation of Condensed Matter, School of Physics, Xi'an Jiaotong University, Xi'an 710049, China
[6] Max Planck Institute for Solid State Research, 70569 Stuttgart, Germany

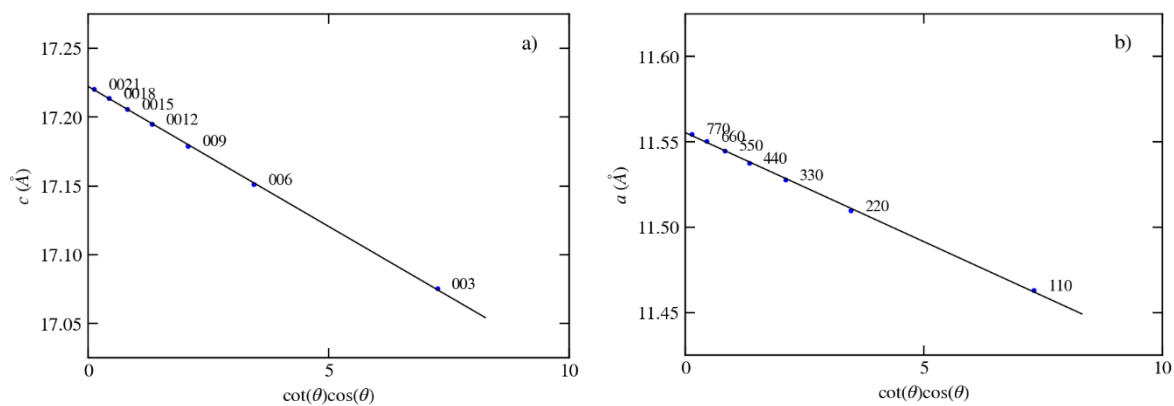

**Fig. S1:** Cohen-Wagner plot for the elimination of sample displacement error. a) For 00l diffraction maxima $c = 17.2223(4)$ Å. b) For hh0 diffraction maxima $a = 11.5554(5)$ Å.

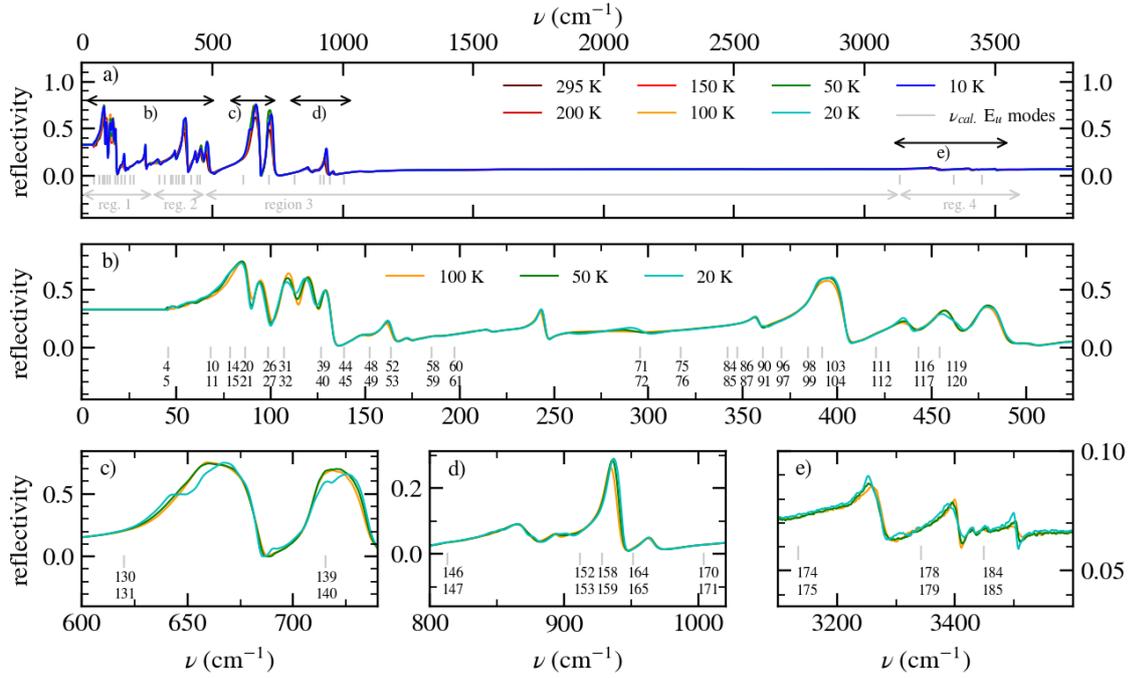

Fig. S2: a) Optical reflectivity ($E \parallel ab$) in far-IR (Sample 1) and mid-IR (Sample 2) regions at various temperatures. b), c), d), and e) panels show selected regions in detail. The vertical grey line segments represent the calculated energy of the phonon modes. The calculated modes are labeled in increasing order as the energy of modes increases. We note that $E_g$, $A_u$, and $A_g$ modes are not shown therefore these numbers are missing below the vertical grey lines. The stars represent the appearance of new phonon modes.

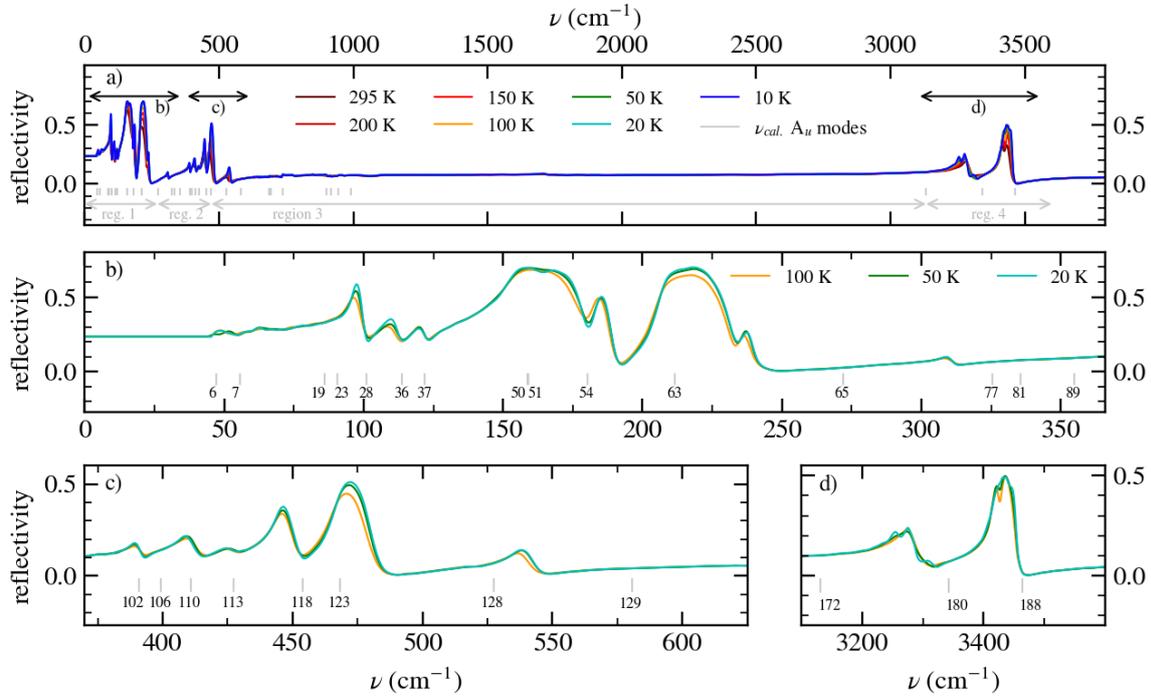

Fig. S3: a) Optical reflectivity ($E \parallel c$) in far-IR (Sample 1) and mid-IR (Sample 2) regions at various temperatures. b), c), d), and e) panels show selected regions in detail. The vertical grey

line segments represent the calculated energy of the phonon modes. The calculated modes are labeled in increasing order as the energy of modes increases. We note that $E_g$, $E_u$, and $A_g$ modes are not shown therefore these numbers are missing below the vertical grey lines.

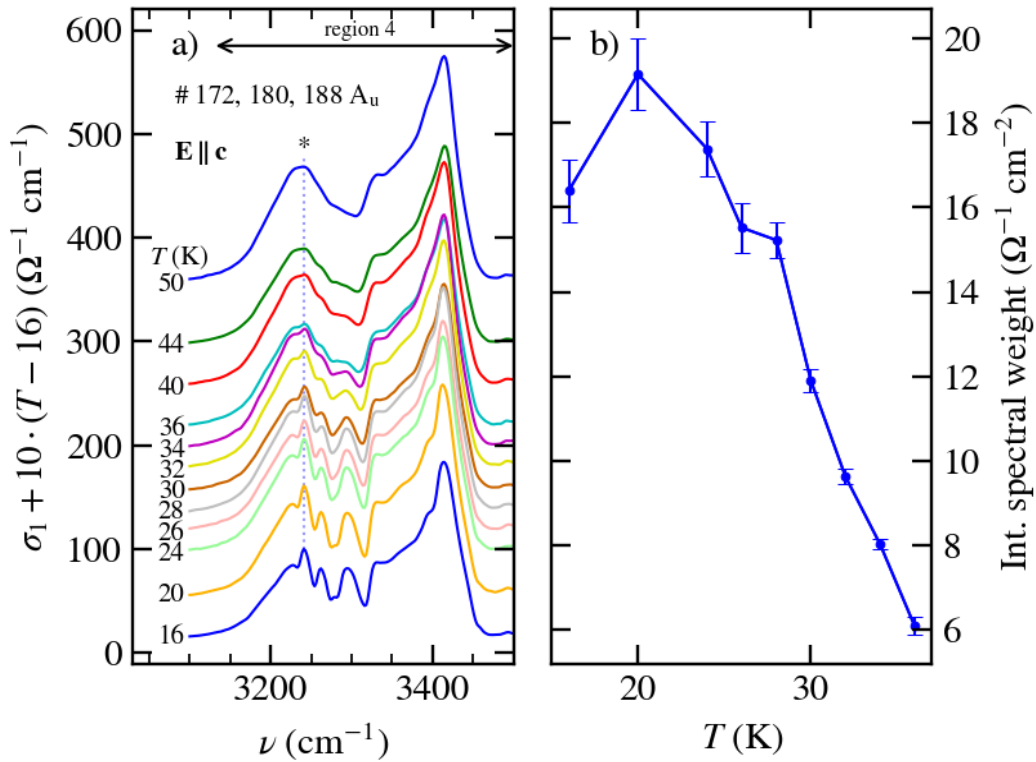

Fig. S4: a) Optical conductivity of Sample 3 in polarization $E \parallel c$ showing the appearance of new phonon modes. b) The integrated spectral weight of one new phonon mode marked by a star in panel a). The numbers of modes represent the modes whose energy should be in the displayed region based on the theoretical calculations.

**Fig. S5:** Atomic displacement in Eu phonon mode # 172, the result of the theoretical calculation. We note that the displacement is enlarged for better visibility.

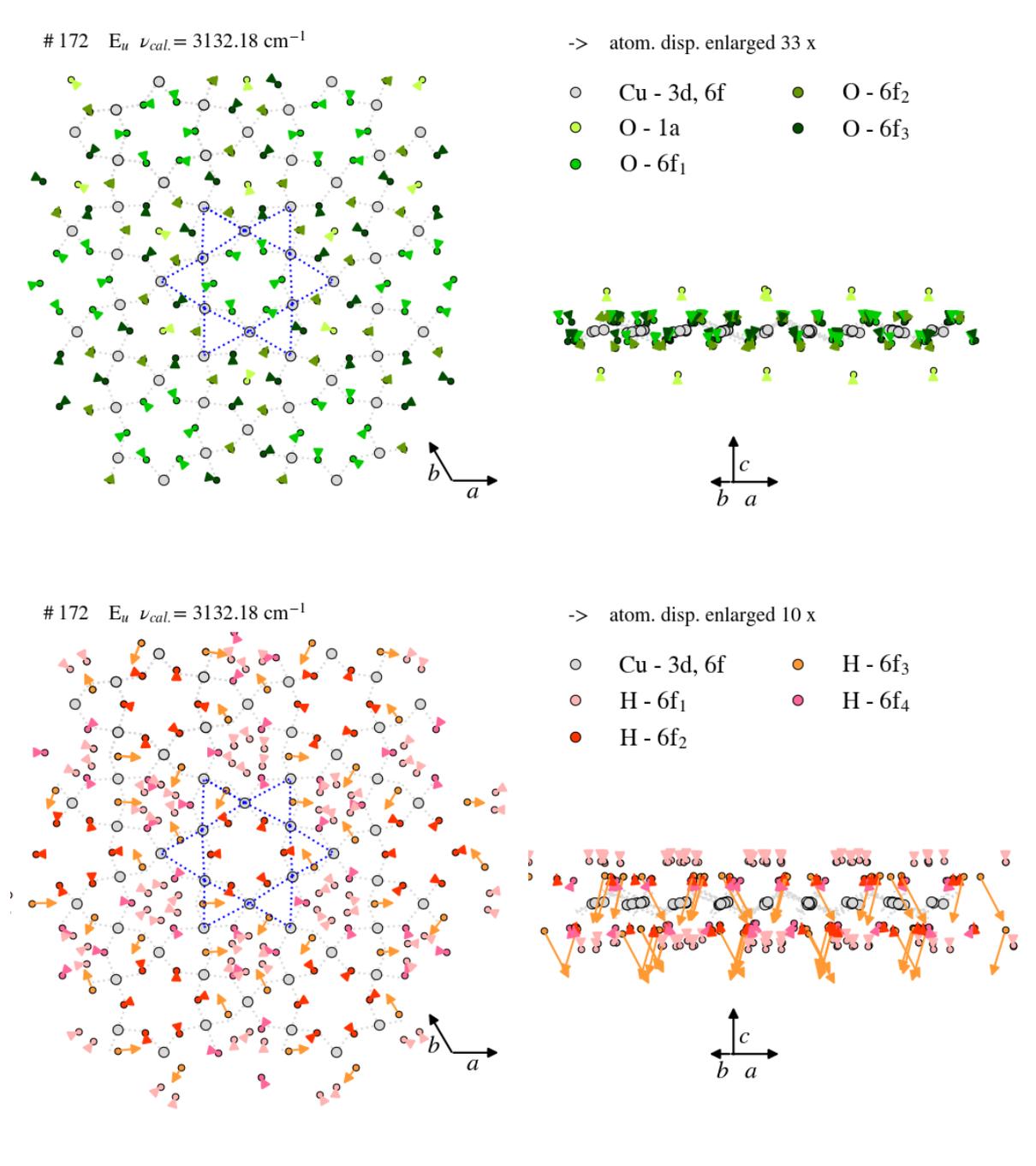

**Fig. S6:** Atomic displacement in Eu phonon mode # 180, the result of the theoretical calculation. We note that the displacement is enlarged for better visibility.

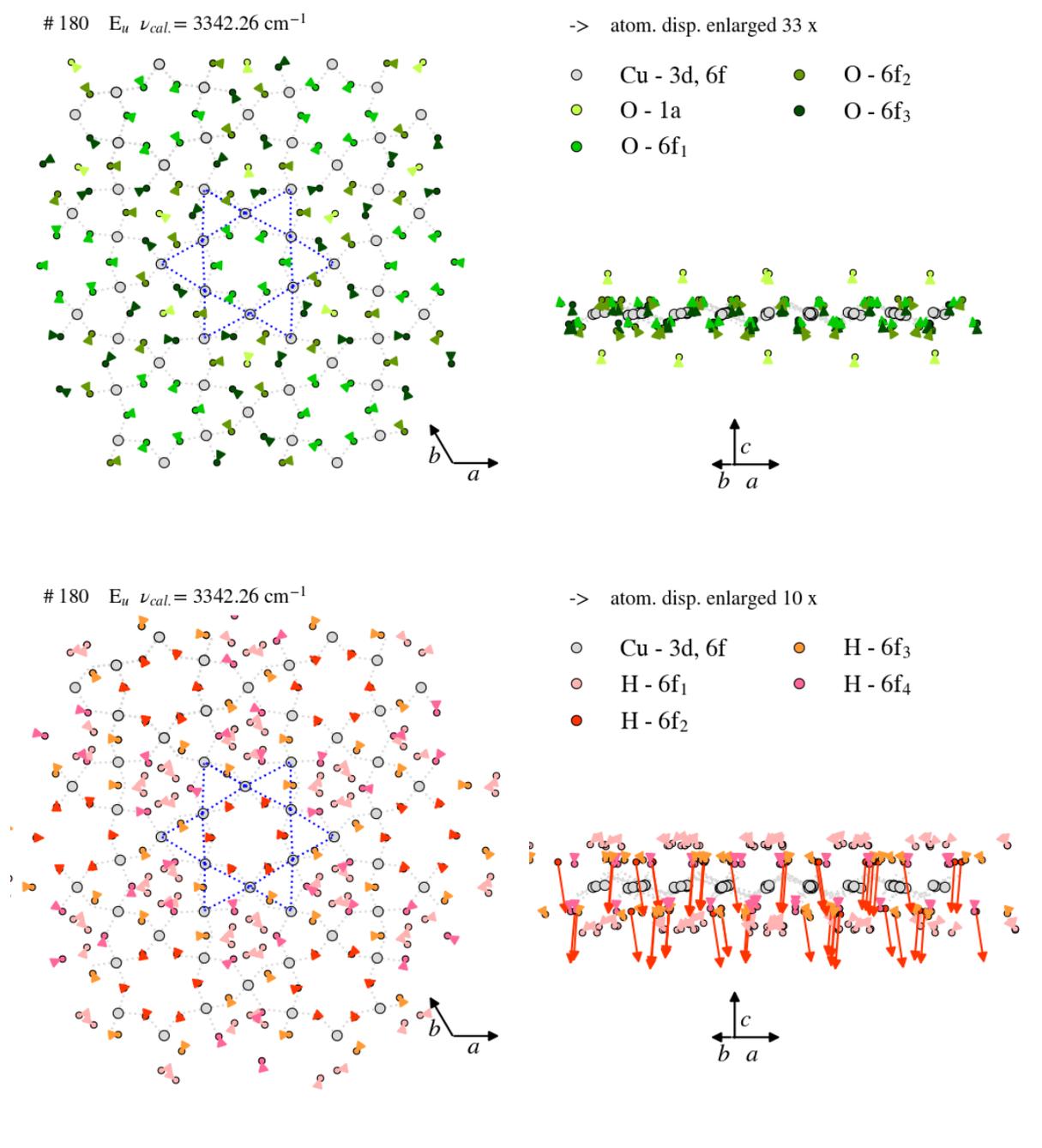

**Fig. S7:** Atomic displacement in Eu phonon mode # 188, the result of the theoretical calculation. We note that the displacement is enlarged for better visibility.

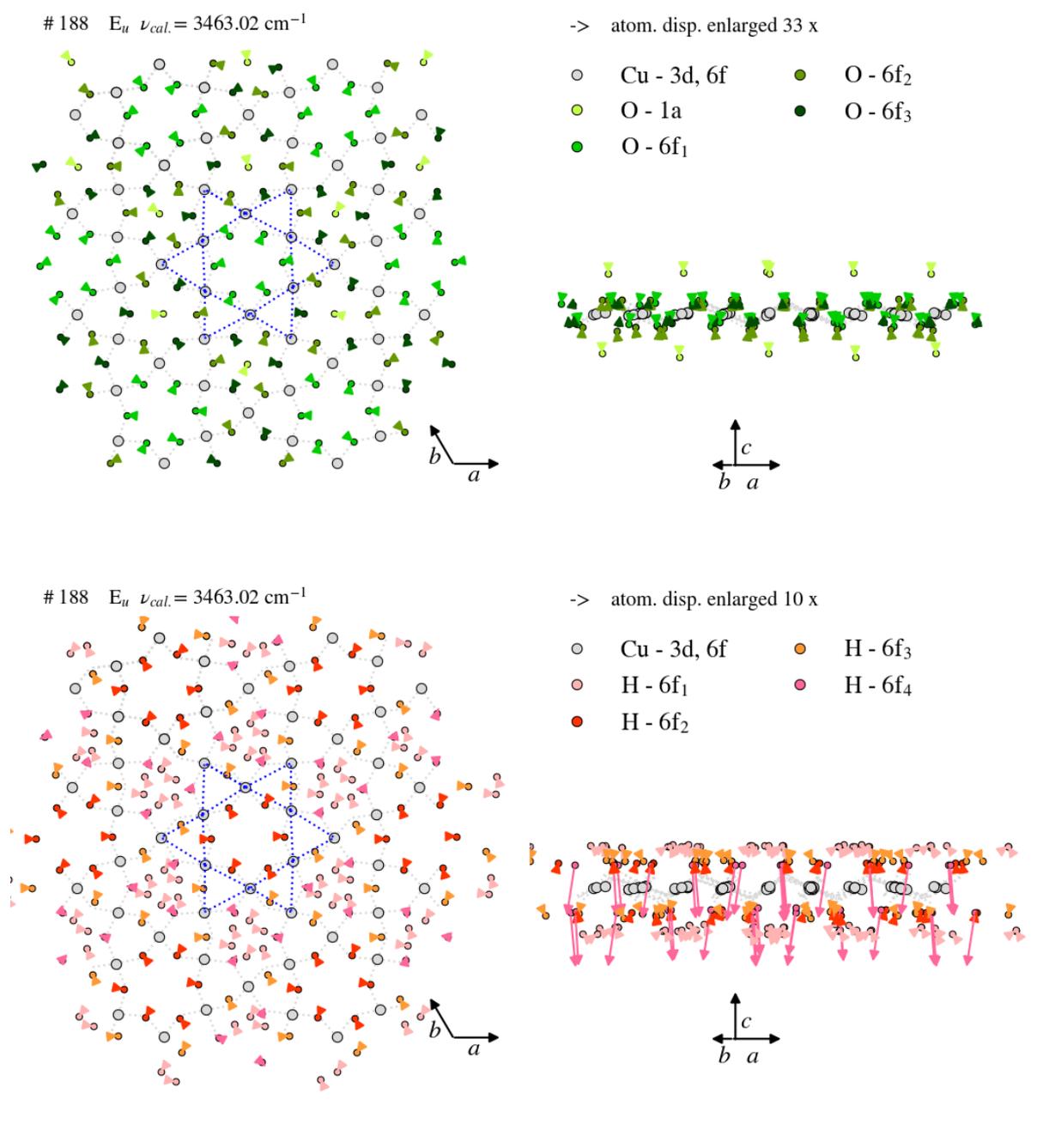

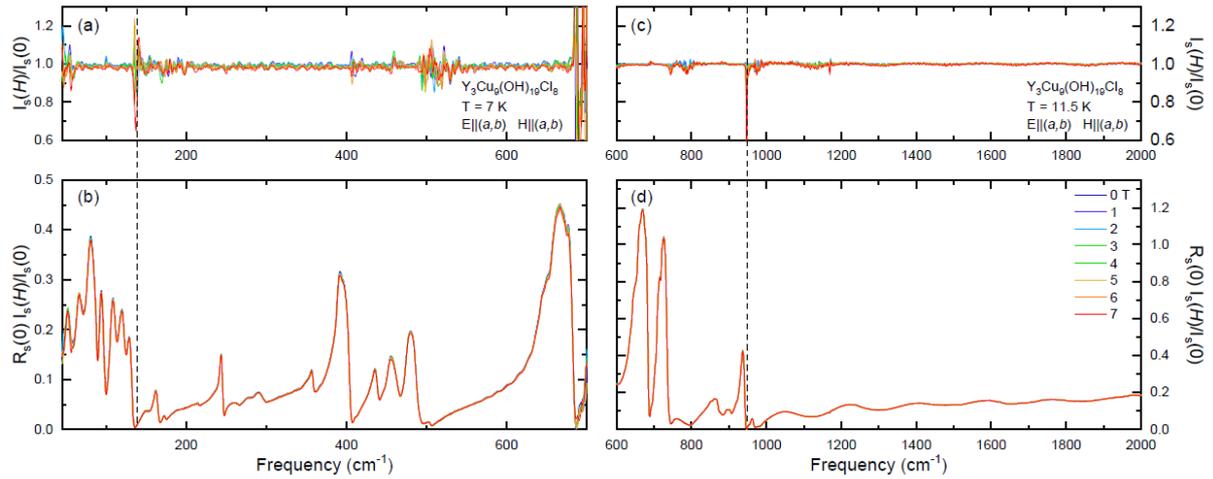

**Fig. S8:** FTIR measurements under external magnetic fields were performed in Voigt geometry, (**E** ∥ **ab**), (**H** ∥ **ab**) using Bruker 113v and a superconducting magnet. a) Relative spectra were recorded under magnetic fields up to 7 T at a temperature of 7 K in the FIR range: $I_s(H)/I_s(0)$ (reflected intensity under field divided by the reflected intensity at zero field). b) Reflectance was obtained by multiplication of the zero-field reflectance $R_s(0)$ (for reference a gold mirror was used) with the relative spectra: $R_s(0) \cdot I_s(H)/I_s(0)$. c,d) Measurements in the MIR range at 11.5 K. For magnetic fields up to 7 T we do not observe significant field dependence. Peaks in the relative spectra under magnetic fields are possibly due to weak signal in the resonance minima (see dashed lines, for instance).

**Fig. S9:** Displacement of Cu, Y, Cl, O and H atoms in mode number 10. The following 5 panels are below.

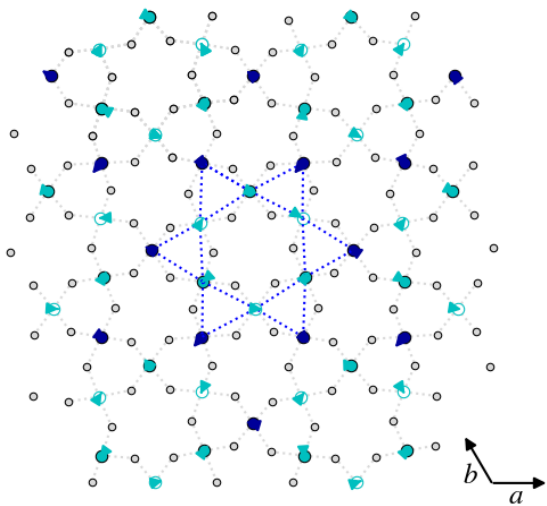

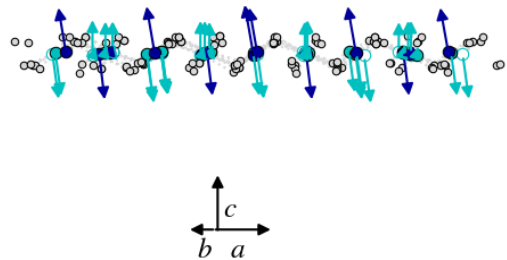

# 10   $E_u$   $\nu_{cal.} = 68.21$ cm$^{-1}$

- Cu - 3d
- Cu - 6f   $z_{Cu\ 6f} > z_{Cu\ 3d}$
- Cu - 6f   $z_{Cu\ 6f} < z_{Cu\ 3d}$
- O - 6f$_1$, 6f$_2$, 6f$_3$
- → atom. disp. enlarged 80 x

\# 10    $E_u$    $\nu_{cal.} = 68.21$ cm$^{-1}$

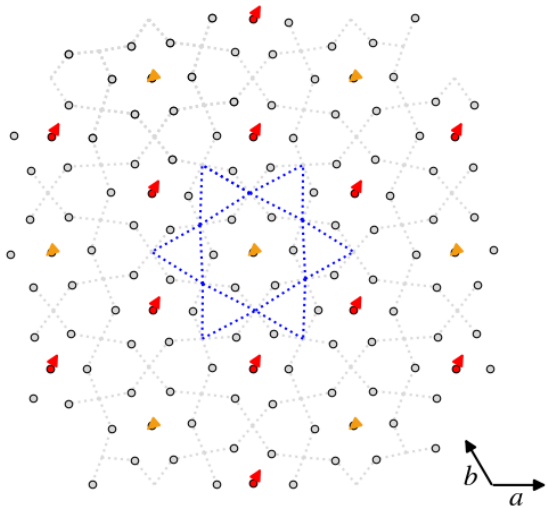
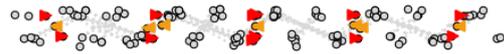

- Y - 2c
- Y - 1b
- O - 6f$_1$, 6f$_2$, 6f$_3$

-> atom. disp. enlarged 80 x

\# 10    $E_u$    $\nu_{cal.} = 68.21$ cm$^{-1}$

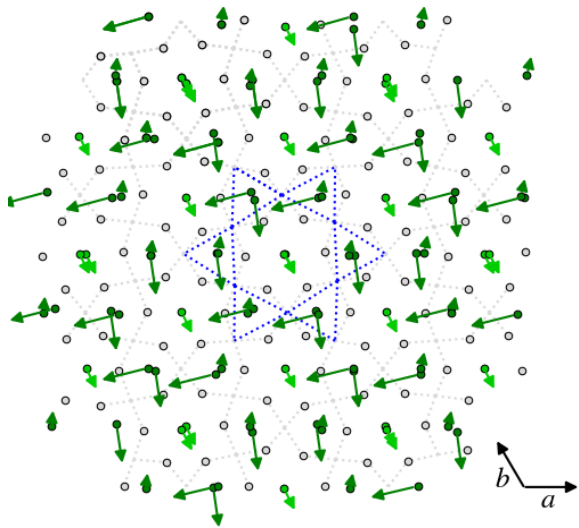
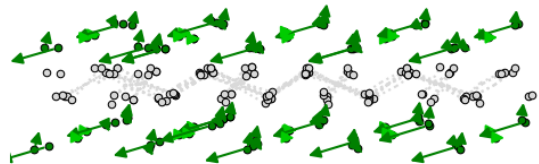

- Cl - 6f
- Cl - 2c
- O - 6f$_1$, 6f$_2$, 6f$_3$

-> atom. disp. enlarged 70 x

#10   $E_u$   $\nu_{cal.} = 68.21$ cm$^{-1}$   -> atom. disp. enlarged 33 x

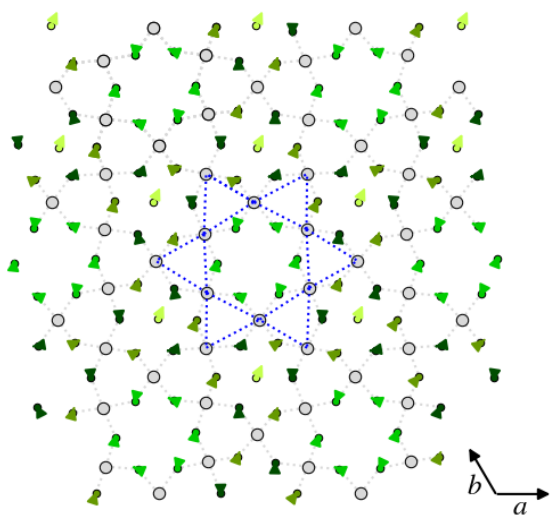
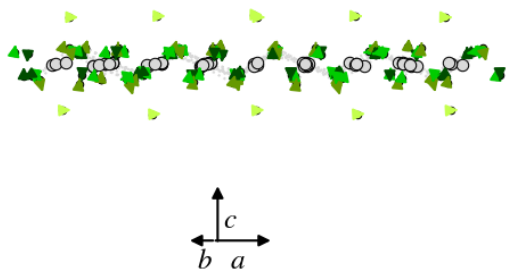

#10   $E_u$   $\nu_{cal.} = 68.21$ cm$^{-1}$   -> atom. disp. enlarged 10 x

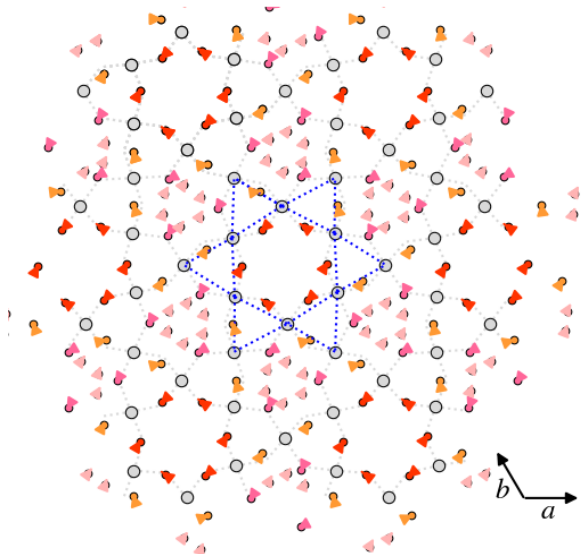
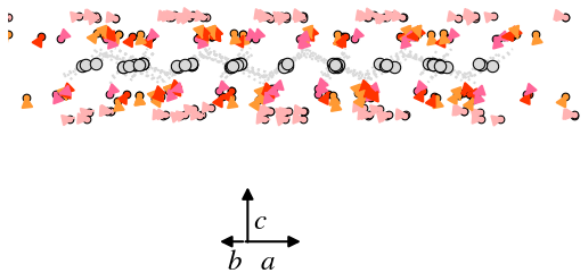

**Fig. S10:** Displacement of Cu, Y, Cl, O and H atoms in mode number 11. The following 5 panels are below.

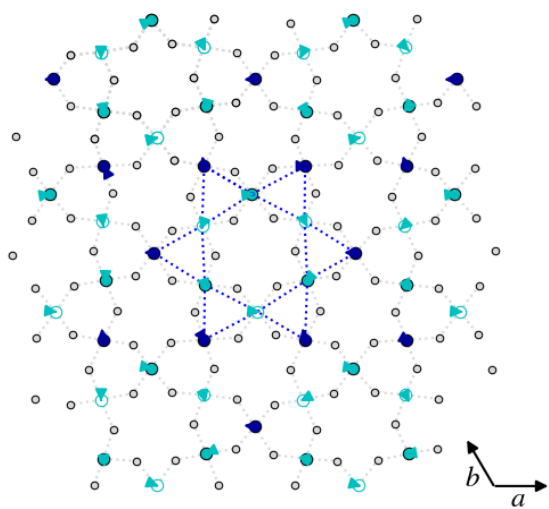
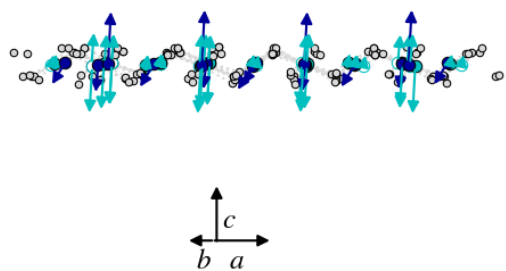

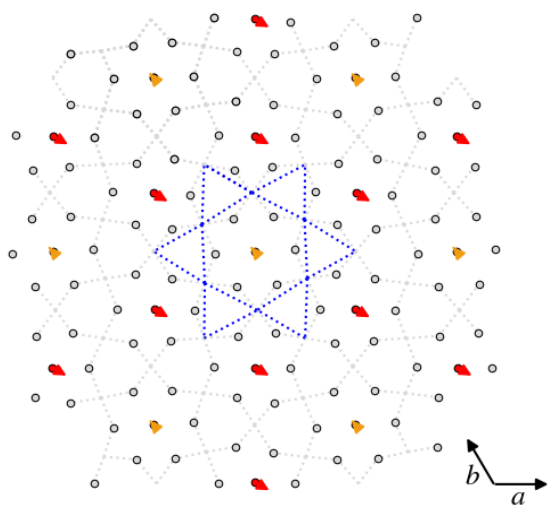
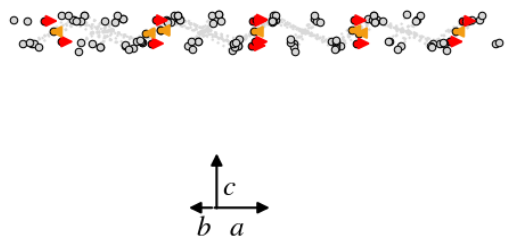

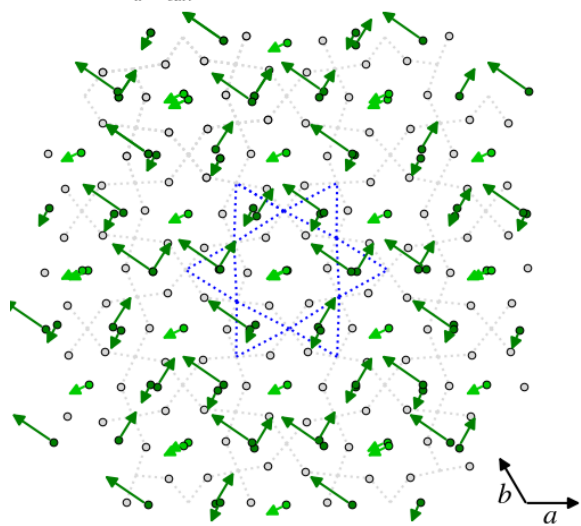
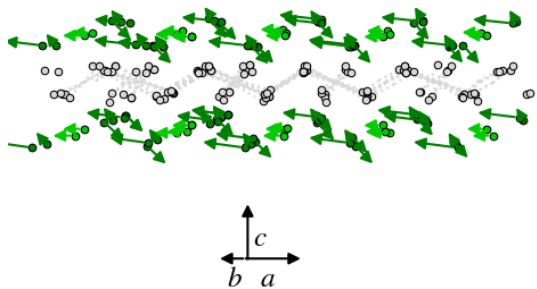

#11   $E_u$   $\nu_{cal.} = 68.21$ cm$^{-1}$

- Cl - 6f
- Cl - 2c
- O - 6f$_1$, 6f$_2$, 6f$_3$
- → atom. disp. enlarged 70 x

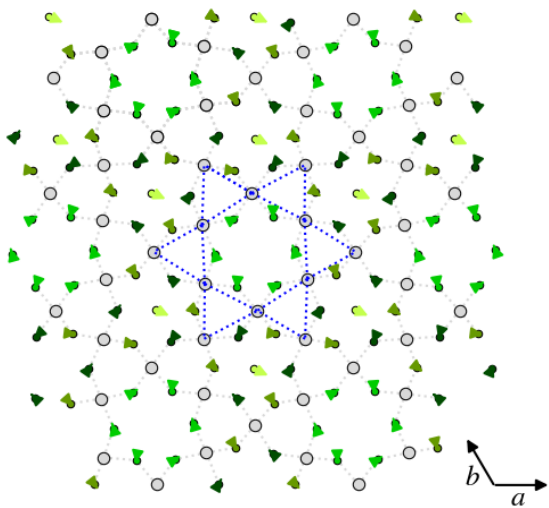
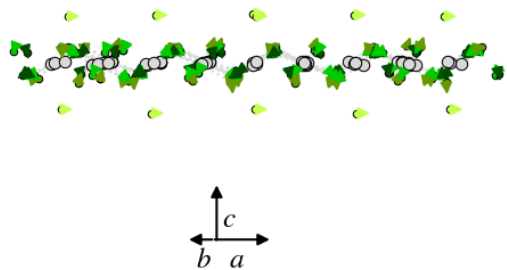

#11   $E_u$   $\nu_{cal.} = 68.21$ cm$^{-1}$

→ atom. disp. enlarged 33 x

- Cu - 3d, 6f
- O - 1a
- O - 6f$_1$
- O - 6f$_2$
- O - 6f$_3$

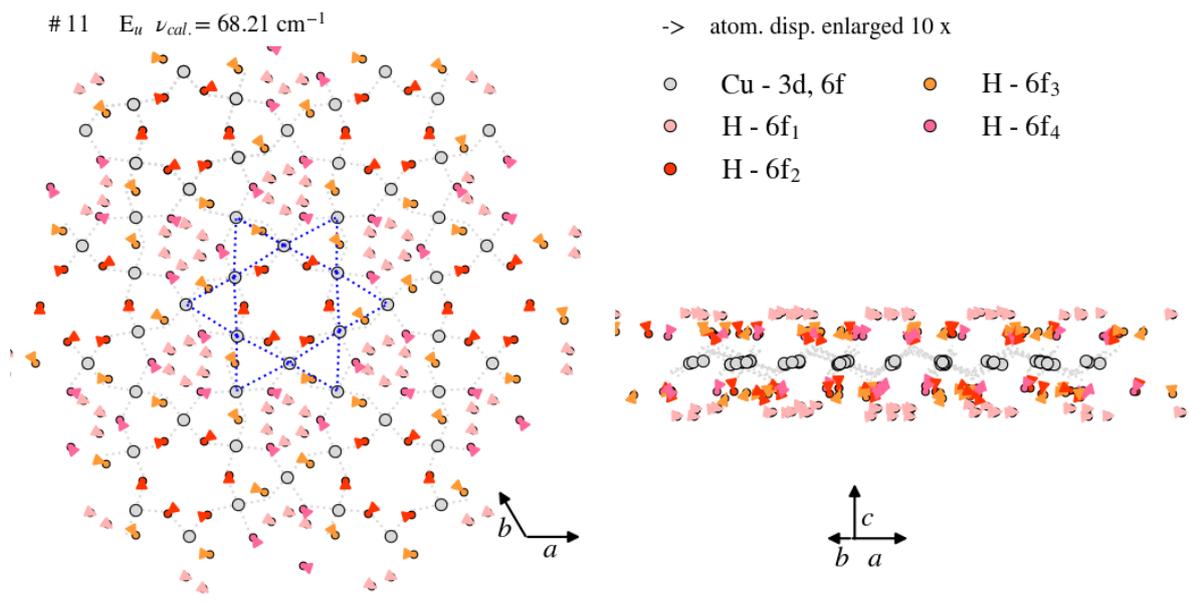

**Fig. S11:** Displacement of Cu, Y, Cl, O and H atoms in mode number 14. The following 5 panels are below.

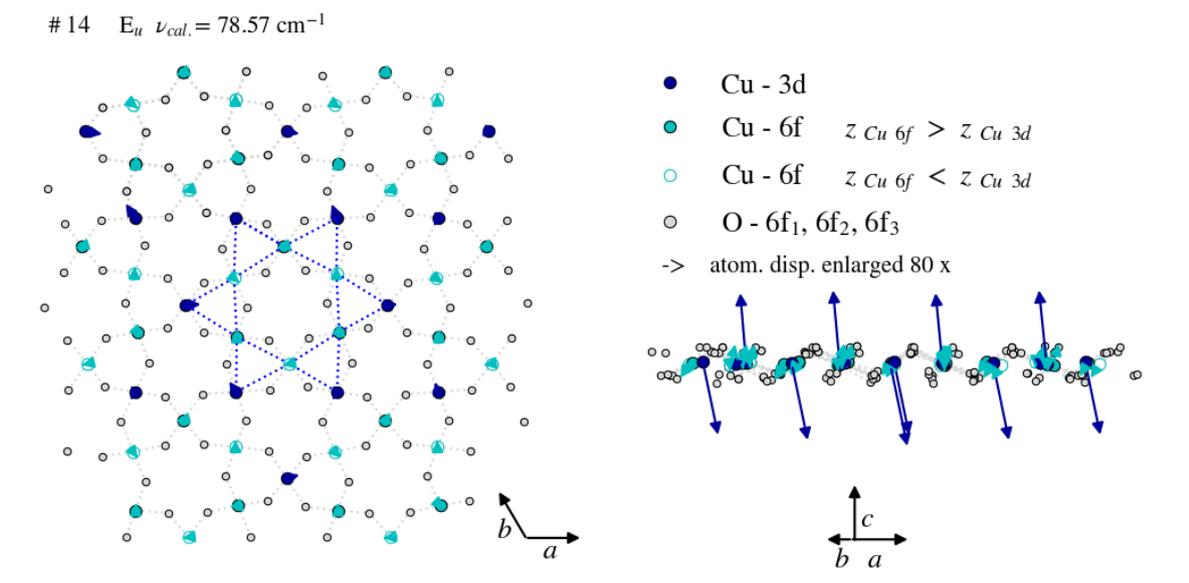

#14  $E_u$  $\nu_{cal.} = 78.57$ cm$^{-1}$

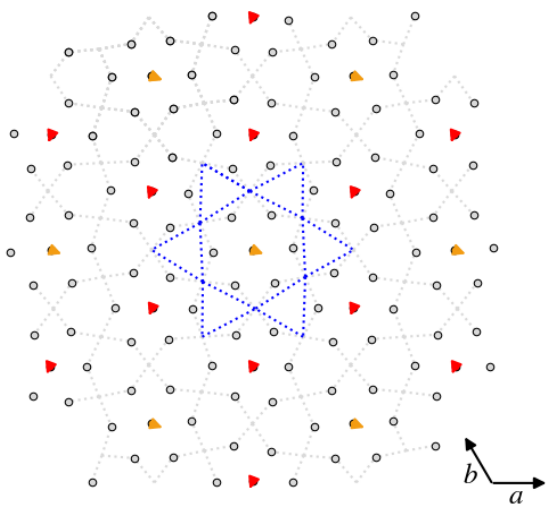

- Y - 2c
- Y - 1b
- O - 6f$_1$, 6f$_2$, 6f$_3$

-> atom. disp. enlarged 80 x

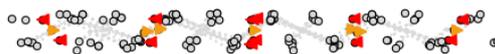

#14  $E_u$  $\nu_{cal.} = 78.57$ cm$^{-1}$

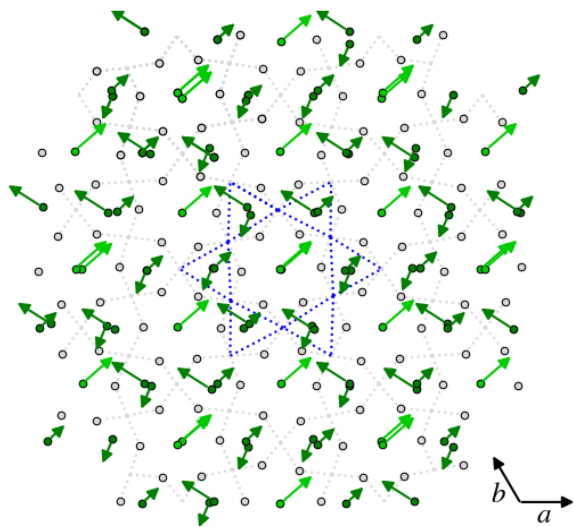

- Cl - 6f
- Cl - 2c
- O - 6f$_1$, 6f$_2$, 6f$_3$

-> atom. disp. enlarged 70 x

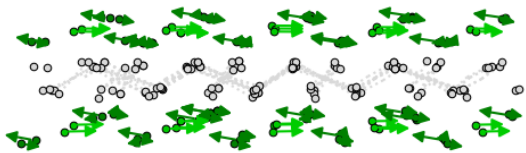

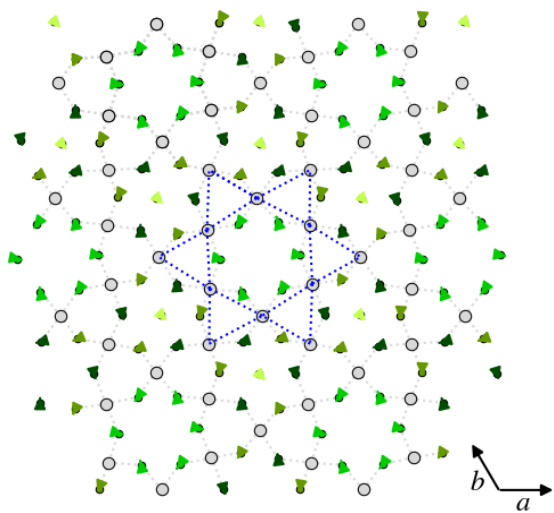
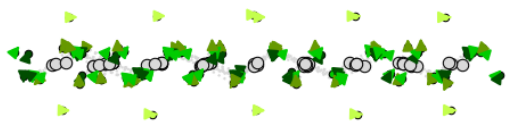

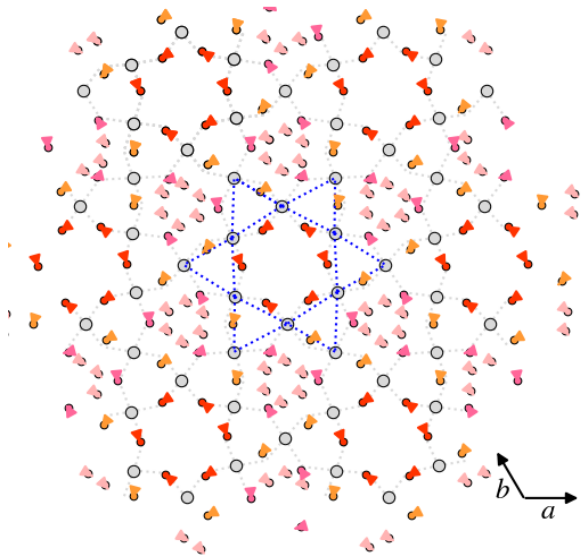
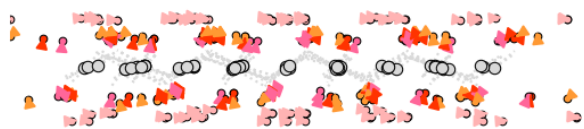

**Fig. S12:** Displacement of Cu, Y, Cl, O, and H atoms in mode number 15. The following 5 panels are below.

#15   $E_u$   $\nu_{cal.} = 78.57$ cm$^{-1}$

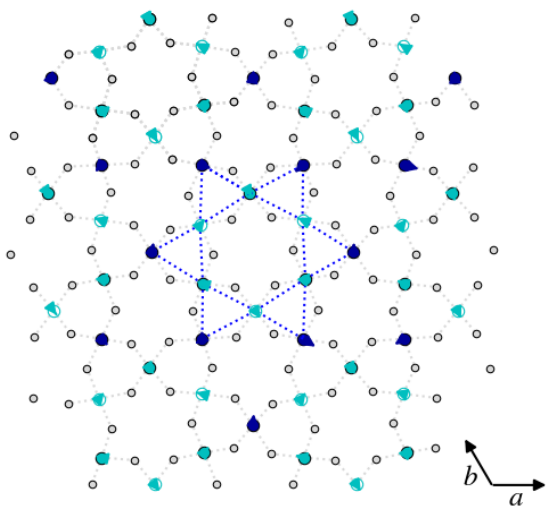
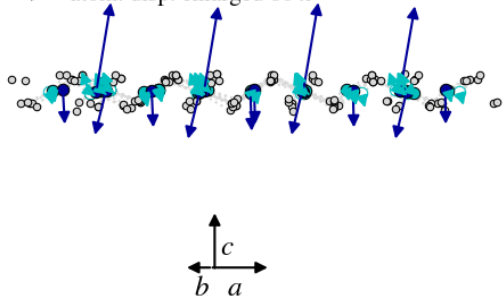

- Cu - 3d
- Cu - 6f   $z_{Cu\ 6f} > z_{Cu\ 3d}$
- Cu - 6f   $z_{Cu\ 6f} < z_{Cu\ 3d}$
- O - 6f$_1$, 6f$_2$, 6f$_3$

-> atom. disp. enlarged 80 x

#15   $E_u$   $\nu_{cal.} = 78.57$ cm$^{-1}$

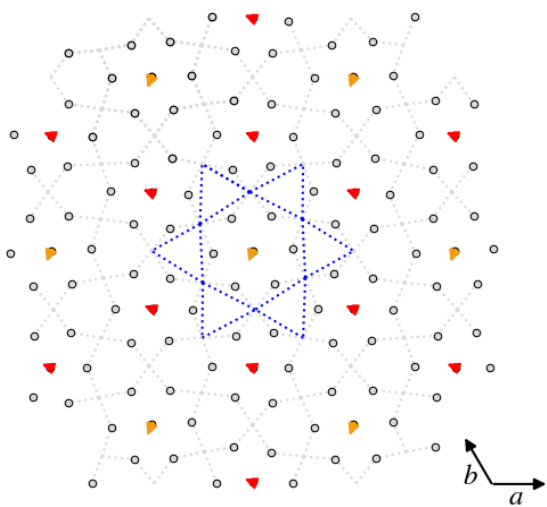
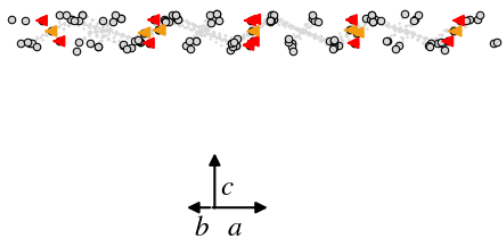

- Y - 2c
- Y - 1b
- O - 6f$_1$, 6f$_2$, 6f$_3$

-> atom. disp. enlarged 80 x

#15  $E_u$  $\nu_{cal.} = 78.57$ cm$^{-1}$

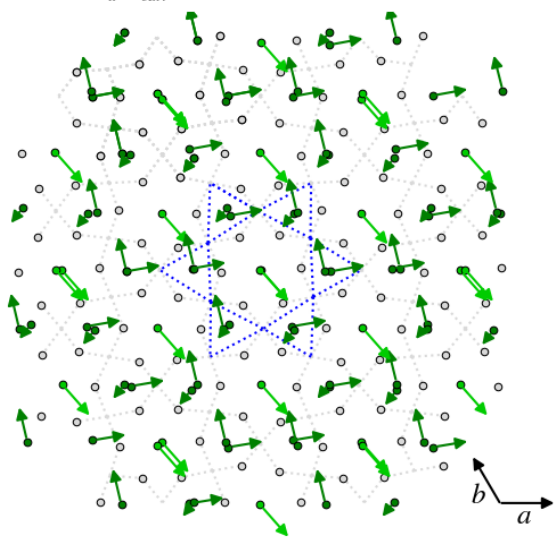
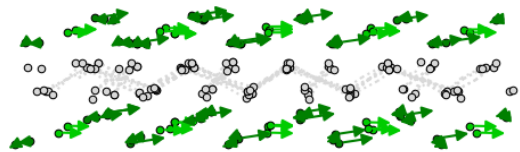

- Cl - 6f
- Cl - 2c
- O - 6f$_1$, 6f$_2$, 6f$_3$
- -> atom. disp. enlarged 70 x

#15  $E_u$  $\nu_{cal.} = 78.57$ cm$^{-1}$

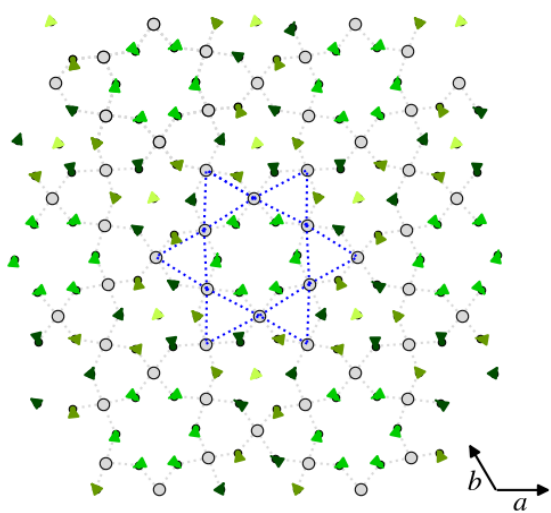
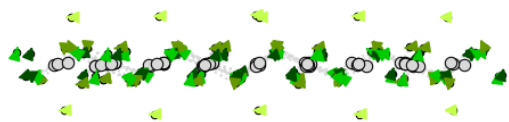

- -> atom. disp. enlarged 33 x
- Cu - 3d, 6f
- O - 1a
- O - 6f$_1$
- O - 6f$_2$
- O - 6f$_3$

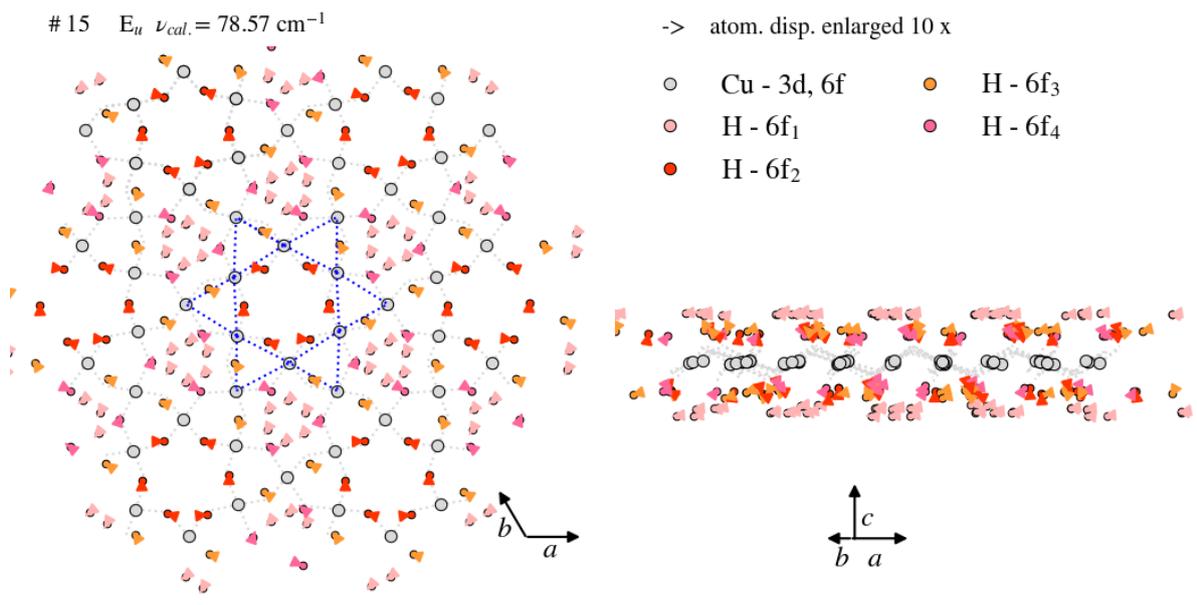

---

**Fig. S13:** Atomic displacement in Eu phonon mode # 131, the result of the theoretical calculation. We note that the displacement is enlarged for better visibility.

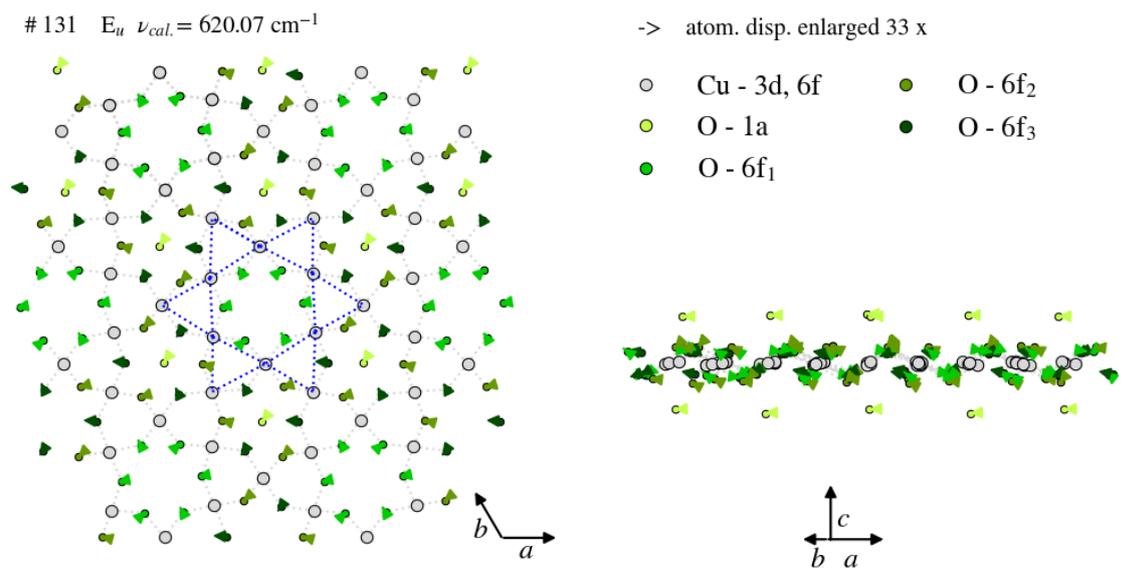

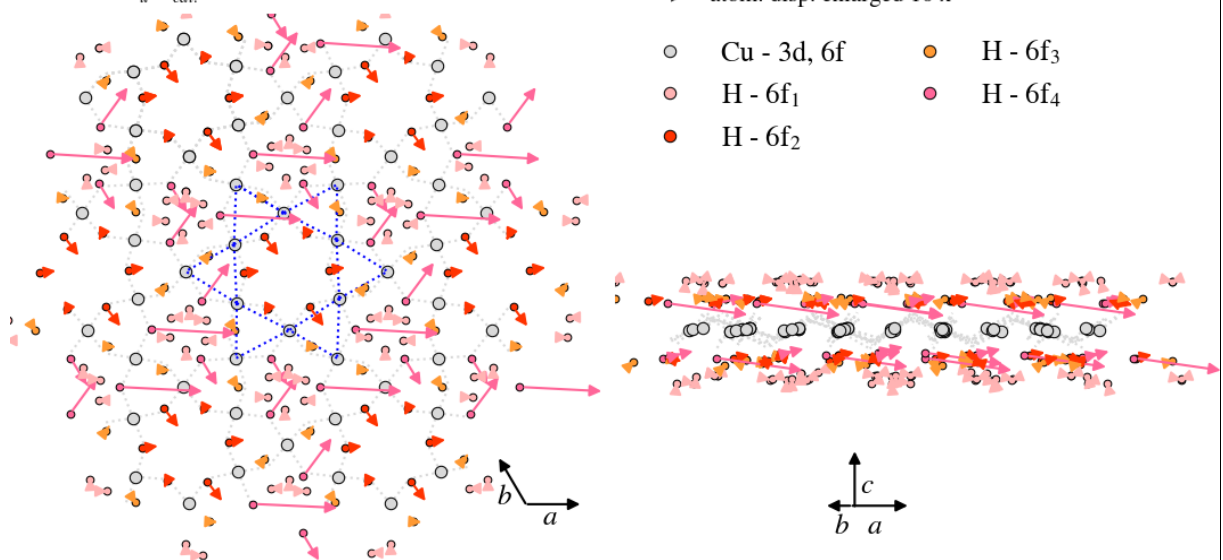